\RequirePackage{xcolor}
\documentclass[11pt]{article}
\usepackage{jcapmod}
\usepackage{tocloft}
\usepackage{ulem}
\usepackage{mathtools}
\usepackage{booktabs}
\usepackage[english]{babel}
\usepackage{amsmath,amssymb,amsbsy,amstext, amsthm, simplewick, amsfonts}
\usepackage{hyperref}
\usepackage{graphicx} 
\usepackage[small]{caption}
\usepackage{slashed}
\usepackage{siunitx}
\usepackage{upgreek}
\usepackage{framed}
\usepackage{wrapfig}
\usepackage{multirow}
\usepackage{bbm}
\usepackage[utf8x]{inputenc}
\usepackage{selinput}
\usepackage{bm}
\usepackage{float}
\usepackage{geometry}
\usepackage{yfonts}
\usepackage{caption}
\usepackage{subcaption}
\usepackage{sidecap}
\usepackage{longtable}
\usepackage{anyfontsize}
\usepackage{dsfont}
\usepackage{tikz}
\usepackage{relsize}
\usepackage{tcolorbox}
\usepackage{braket}
\usepackage{mathtools}

\usepackage[framemethod=default]{mdframed}
\newmdenv[skipabove=7pt,
skipbelow=7pt,
rightline=false,
leftline=false,
topline=false,
bottomline=false,
backgroundcolor=gray!10,
linecolor=gray,
innerleftmargin=5pt,
innerrightmargin=5pt,
innertopmargin=5pt,
innerbottommargin=5pt,
leftmargin=0cm,
rightmargin=0cm,
linewidth=4pt]{eBox}
\newmdenv[skipabove=7pt,
skipbelow=7pt,
rightline=false,
leftline=false,
topline=false,
bottomline=false,
backgroundcolor=gray!10,
linecolor=gray,
innerleftmargin=5pt,
innerrightmargin=5pt,
innertopmargin=-5pt,
innerbottommargin=5pt,
leftmargin=0cm,
rightmargin=0cm,
linewidth=4pt]{eBox2}
\usepackage[framemethod=default]{mdframed}
\newmdenv[skipabove=7pt,
skipbelow=7pt,
rightline=true,
leftline=true,
topline=true,
bottomline=true,
backgroundcolor=gray!15,
linecolor=gray,
innerleftmargin=5pt,
innerrightmargin=5pt,
innertopmargin=5pt,
innerbottommargin=5pt,
leftmargin=0cm,
rightmargin=0cm,
linewidth=0.75pt]{eBox3}

\usepackage{colortbl}
\definecolor{greyish2}{rgb}{.96,.96,.96}

\makeatletter
\newlength{\apb@width}
\newcommand{\autoparbox}[2][c]{\settowidth{\apb@width}{#2}\parbox[#1]{\apb@width}{#2}}

\makeatother


\def\beq{\begin{equation}}
\def\eeq{\end{equation}}

\newcommand{\dd}{\mathrm{d}} 
\newcommand{\Psig}{P}

\usetikzlibrary{arrows,calc,decorations.pathmorphing,decorations.markings,shapes.misc}
\pgfkeys{/tikz/.cd,
  circle color/.initial=black,
  circle color/.get=\circlecolor,
  circle color/.store in=\circlecolor,
}
\tikzset{/pgf/decoration/.cd,
    number of sines/.initial=10,
    angle step/.initial=20,
}
\newdimen\tmpdimen\pgfdeclaredecoration{complete sines}{initial}
{
    \state{initial}[
        width=+0pt,
        next state=move,
        persistent precomputation={
            \pgfmathparse{\pgfkeysvalueof{/pgf/decoration/angle step}}%
            \let\anglestep=\pgfmathresult%
            \let\currentangle=\pgfmathresult%
            \pgfmathsetlengthmacro{\pointsperanglestep}%
                {(\pgfdecoratedremainingdistance/\pgfkeysvalueof{/pgf/decoration/number of sines})/360*(\anglestep)}%
        }] {}
    \state{move}[width=+\pointsperanglestep, next state=draw]{
        \pgfpathmoveto{\pgfpointorigin}
    }
    \state{draw}[width=+\pointsperanglestep, switch if less than=1.25*\pointsperanglestep to final, 
        persistent postcomputation={
        \pgfmathparse{mod(\currentangle+\anglestep, 360)}%
        \let\currentangle=\pgfmathresult%
    }]{%
        \pgfmathsin{+\currentangle}%
        \tmpdimen=\pgfdecorationsegmentamplitude%
        \tmpdimen=\pgfmathresult\tmpdimen%
        \divide\tmpdimen by2\relax%
        \pgfpathlineto{\pgfqpoint{0pt}{\tmpdimen}}%
    }
    \state{final}{
        \ifdim\pgfdecoratedremainingdistance>0pt\relax
            \pgfpathlineto{\pgfpointdecoratedpathlast}
        \fi
   }
}

\allowdisplaybreaks[1]
\begin{document}


\newgeometry{top=3cm, bottom=3cm, left=3cm, right=3cm}

\begin{titlepage}
\setcounter{page}{1} \baselineskip=15.5pt 
\thispagestyle{empty}

\begin{center}
{\fontsize{18}{18} \bf No-go Theorem for Scalar-Trispectrum-Induced Gravitational Waves}\\[14pt]
\end{center}

\vskip 20pt
\begin{center}
\noindent
{\fontsize{12}{18}\selectfont 
Sebastian Garcia-Saenz,$^{1}$  Lucas Pinol,$^{2}$
Sébastien Renaux-Petel,$^{3}$ and 
Denis Werth$^{3}$\hskip 1pt}
\end{center}

\begin{center}
    \vskip8pt
\textit{$^1$ Department of Physics, Southern University of Science and Technology, \\Shenzhen 518055, China}

\vskip 8pt
\textit{$^2$ Instituto de Física Teórica UAM/CSIC, U. Autónoma de Madrid, C/ Nicolás Cabrera 13-15, Cantoblanco, 28049, Madrid, Spain}

\vskip 8pt
\textit{$^3$ Sorbonne Université, CNRS, UMR 7095, Institut d’Astrophysique de Paris,\\ 98 bis bd Arago, 75014 Paris, France}
\end{center}

\vspace{0.4cm}
\begin{center}{\bf Abstract}
\end{center}
\noindent
We show that the contribution of the primordial trispectrum to the energy density of the scalar-induced stochastic gravitational wave background cannot exceed the one from the scalar power spectrum in conventional inflationary scenarios.
Specifically, we prove in the context of scale-invariant theories that neither regular trispectrum shapes peaking in so-called equilateral configurations, nor local trispectrum shapes diverging in soft momentum limits, can contribute significantly.
Indeed, those contributions are always bound to be smaller than an order-one (or smaller) number multiplying the relative one-loop correction to the scalar power spectrum, necessarily much smaller than unity in order for the theory to be under perturbative control.
Since a no-go theorem is only worth its assumptions, we also briefly discuss a toy model for a scale-dependent scalar spectrum, which confirms the robustness of our no-go result.
\end{titlepage}
\restoregeometry

\newpage
\setcounter{tocdepth}{3}
\setcounter{page}{2}

\linespread{1.6}
\tableofcontents
\linespread{1.1}

\newpage

\section{Introduction}
\label{sec:intro}

Gravitational waves (GWs) of primordial origin are one of
the most important probes of the early universe, allowing us to experimentally access---if they are ever detected---the earliest events of the hot Big Bang history, and possibly even further back into a putative inflationary phase. While vacuum GWs of primordial quantum origin provide an interesting but likely minute signal, it has been appreciated in recent years that GWs sourced by scalar cosmological perturbations are potentially significant and perhaps directly observable in forthcoming GW experiments (see e.g.~\cite{Bartolo:2016ami,Domenech:2021ztg,Achucarro:2022qrl,LISACosmologyWorkingGroup:2022jok} for reviews).

\vskip 4pt
In the context of inflation, scalar perturbations---most conveniently quantified by the curvature perturbation $\zeta$---have an amplitude that is again woefully small on the length scales measured in the cosmic microwave background (CMB) and with large scale structure surveys. This corresponds however to physics occurring 50-60 e-folds before the end of inflation, leaving a comparatively huge time span afterward which is essentially unconstrained by observations. This unknown window is of obvious importance, mainly as a basic probe of inflation on small scales but also because of other critical effects such as the formation of primordial black holes.

\vskip 4pt
This has motivated theorists to explore motivated scenarios beyond ``vanilla'' ones in which scalar perturbations are in some way enhanced at small scales beyond what the extrapolation of the CMB-scale amplitude would predict (see references in the above reviews). Among these, most studies have concentrated on the Gaussian approximation, i.e.\ considering that the primordial curvature perturbation follows a Gaussian statistics, entertaining the possibility that the scalar power spectrum may also have strong features on small scales, thus leading to a GW signal that is interesting both in its amplitude and frequency dependence. However, 
the energy density of the scalar-induced stochastic gravitational wave background (SGWB) depends on the four-point correlation function of $\zeta$ and therefore includes in general a non-Gaussian (NG) component, set by the connected four-point function of the primordial curvature perturbation, also known as the trispectrum.  This component is the object of this work.

\vskip 4pt
In this respect, we deem it useful to clarify some nomenclature used in the literature.
So far, the consequences of scalar perturbations not following Gaussian statistics on the scalar-induced SGWB have only been studied by considering a local ansatz for the primordial curvature perturbation, of the type $\zeta=\zeta_\textrm{G}+f_\textrm{NL} \zeta_\textrm{G}^2$, where $\zeta_\mathrm{G}$ is a Gaussian variable.
This parameterisation of the observable curvature perturbation $\zeta$ has two implications.
First, its Gaussian statistics (encoded in the power spectrum) are modified, as e.g.\ the power spectrum $P_\zeta=P_{\zeta_\mathrm{G}}+3 f_\mathrm{NL}^2 P_{\zeta_\mathrm{G}}^2$.
Second, it displays non-Gaussianities (encoded in higher-order connected correlation functions), as e.g.\ the connected four point function
$\braket{\zeta^4}-3\braket{\zeta^2}^2$, which includes terms like $f_\mathrm{NL}^2 P_{\zeta_\mathrm{G}}^3$, is non-zero.

\vskip 4pt
Several works claim to have considered the impact of primordial non-Gaussianities on the scalar-induced SGWB, but were applying the Gaussian approximation in which the SGWB energy density is solely determined by the primordial power spectrum \cite{Cai:2018dig,Cai:2019amo,Ragavendra:2020sop,Yuan:2020iwf}.
As such, this does not correspond to effects \textit{due} to primordial non-Gaussianities, but rather to corrections to the Gaussian statistics encoded in $P_\zeta$ from non-linearities, which is anyway the only observable quantity as we have no access to the putative tree-level signal $P_{\zeta_\mathrm{G}}$.
Therefore, the effects considered in these works correspond to the first of the two consequences of the local ansatz that we have just described, and are completely distinct from the ones of interest here: primordial non-Gaussianities.

\vskip 4pt
On the other hand, a few interesting studies have genuinely taken into account the impact of the trispectrum on the scalar-induced SGWB, having uncovered potentially large effects \cite{Unal:2018yaa,Atal:2021jyo,Adshead:2021hnm}, including a frequency profile that is distinct from the Gaussian counterpart (see also \cite{Garcia-Bellido:2017aan} in a different context).
Yet these latter works also assume a phenomenological local ansatz, together with a small-scale enhancement of the power spectrum of $\zeta_\textrm{G}$ in a strongly scale-dependent way, and a large or even dominant non-Gaussian effect in the sense of having $f_\textrm{NL} {\cal P}^{1/2}_{\zeta_\textrm{G}}$ of order one or larger.
Of course, the unknown physics of the primordial universe may well possibly result in such situations, where the local ansatz is exact and not meant to emerge from a perturbative series expansion, but the subject clearly deserves further investigation. For instance, the primordial phenomena leading to a substantial enhancement of the primordial power spectrum on small scales likely results in general in trispectrum signals significantly more complex than the simple one captured by the local ansatz. Moreover, in concrete models of the early universe, theoretical control over the computations leading to the predictions for the statistical properties of $\zeta$ necessarily results in bounds on the amplitude of non-Gaussian effects.

\vskip 4pt
This leads us to the main question that we want to address in this work: \textit{can the trispectrum be observationally relevant for the scalar-induced SGWB in conventional set-ups where NGs provide subleading corrections to the Gaussian scalar signal?} Our meaning of ``subleading'' is very precise, since we ask that NGs be generated by interactions consistent with weak coupling, i.e.\ that they maintain perturbative control. An additional novelty of our work is that we do not restrict our attention to local NGs as was done in previous studies, as we also investigate different shapes or momentum dependencies, in relationship with the various types of inflationary physics that give rise to them. Formulating a precise quantitative criterion ensuring perturbative control in strongly scale-dependent theories is difficult and model-dependent. Hence, as a first concrete step, our focus in this paper will be on scale-invariant scenarios, with the investigation of some scale-dependent effects only through a toy model as a way to check the robustness of our results.

\vskip 4pt
We consider first trispectrum shapes that peak in generic ``equilateral'' momentum configurations, specifically the trispectrum predicted by the simplest low-energy effective field theory of single-field inflationary fluctuations~\cite{Creminelli:2006xe, Cheung:2007st}, which is computationally equivalent to $P(X,\phi)$ models of inflation characterised by a non-standard kinetic structure.
Using a semi-analytical approach, we confirm the expectation that the condition of perturbative control over the theory leads to a suppressed trispectrum-induced GW signal.
Specifically, the requirement that the one-loop correction to the power spectrum be subdominant relative to the tree-level one bounds the energy scale $H$ of inflation to be much smaller than a strong coupling scale $\Lambda_\star$.
We find the contribution from the connected four-point function, relative to the disconnected one, to be of order $\mathcal{O}(10^{-1})\left(H/\Lambda_\star\right)^4$.

\vskip 4pt
We then investigate the case of a trispectrum that arises from the non-linear super-Hubble effects of additional light scalar fields beyond $\zeta$, and that consists in the sum of two shapes whose amplitudes are parameterised by the numbers $\tau_\mathrm{NL}$ and $g_{\mathrm{NL}}$: the so-called local trispectrum shapes (see Ref.~\cite{Wands:2010af} for a review).
First, we show that due to the symmetry properties of the integrals in the trispectrum-induced tensor power spectrum, the $g_\mathrm{NL}$-shape does not contribute at all.
As for the $\tau_\mathrm{NL}$-shape, which can be interpreted microscopically as resulting from interactions due to the exchange of massless particles on super-horizon scales, we investigate separately the contributions from the different $s$, $t$ and $u$ channels corresponding to different exchanged momenta.
The contribution from the $s$-channel vanishes for the same reason than the $g_\mathrm{NL}$ shape does, and we compute analytically the contributions from $t$ and $u$ channels, which we show to be dominated by the so-called collapsed limit of the trispectrum shape, corresponding to the exchange of particles with soft momenta.
Since the collapsed limit of correlation functions is known to be a robust probe of the inflationary field content (see, e.g.~the review \cite{Achucarro:2022qrl}), specifically of the masses and spins of the fields whose fluctuations contribute to the correlators of $\zeta$, this result motivates the analysis of more general collapsed trispectra that take into account the mass and spin of the exchanged particle.
We therefore also compute the trispectrum-induced GW energy density from massive and spinning particles during inflation, and show that it is always smaller than the one from the local trispectrum and, as expected, more and more suppressed as the mass of the exchanged particle gets larger.
But importantly, we prove completely generally that the contribution from the local $\tau_\mathrm{NL}$-shape relative to the Gaussian one is always less than $4\,P^{\rm(1-loop)}_{\zeta}/P^{\rm(tree)}_{\zeta} $, with equality in single-field inflation only.
Therefore, again, the requirement of perturbative control over the theory consisting in asking that loop corrections be suppressed, puts a severe bound on the amplitude of the local-trispectrum-induced gravitational waves in scale-invariant theories.

\vskip 4pt
We then finish with some conclusive thoughts and we discuss possible extensions; in particular we present first results for a simple toy model of a scale-dependent power spectrum, for which we show that our no-go theorem is upheld.

\vskip4pt
\paragraph{Notations and conventions.}
We use natural units, $\hbar = c \equiv 1$.
We use Greek letters for spacetime indices, $\mu =0,1,2,3$, and Latin letters for spatial indices, $i=1,2,3$. Physical time is $t$ and conformal time is $\eta$.
Spatial vectors are denoted by $\bm x$. The corresponding three-momenta are $\bm k$, with magnitudes $k = |\bm k|$ and unit vectors written as $\hat{\bm{k}} = \bm k/k$.
We use Latin letters from the beginning of the alphabet to label the momenta of the different legs of a correlation function, i.e.~$\bm k_a$ is the momentum of the $a$-th leg. A prime on a correlator is defined to mean that we drop the momentum conserving delta function:
\begin{equation*}
    \braket{\mathcal{O}_{\bm{k}_1} \ldots \mathcal{O}_{\bm{k}_4}} = (2\pi)^{3}\,\delta(\bm{k}_1 + \ldots + \bm{k}_4)\braket{\mathcal{O}_{\bm{k}_1} \ldots \mathcal{O}_{\bm{k}_4}}' \,.
\end{equation*}

\newpage
\section{Gravitational waves induced by the scalar four-point function}
\label{sec:preliminaries}

The object of central interest is the energy density of the stochastic gravitational wave background non-linearly sourced by scalar fluctuations when they re-enter the Hubble radius after inflation. In this section, after a lightning introduction to the subject, we review the derivation of the master integral for the induced GWs and briefly discuss its main properties, particularly the ones that will play an important role in the following. More details can be found in the review \cite{Domenech:2021ztg}.

\subsection{Tensor perturbations}

The aim is to solve the equation of motion for the tensor perturbation $h_{ij}$, which we decompose in Fourier modes and polarizations as
\begin{equation}
    h_{ij}(\eta, \bm{x}) = \sum_{\lambda = +, \times}\int \frac{\mathrm{d}^3\bm{k}}{(2\pi)^3}\, \mathrm{e}_{ij}^{\lambda}(\hat{\bm{k}})\, h^{\lambda}_{\bm{k}}(\eta) e^{i \boldsymbol{k} \cdot \boldsymbol{x}} \,,
\end{equation}
where the polarization tensors $\mathrm{e}_{ij}^{\lambda}$ are transverse and traceless.\footnote{Different conventions for the normalisation of the tensor perturbation $h_{ij}$ and of the polarization tensors $\mathrm{e}_{ij}^{\lambda}$ lead to different numerical factors in the relationship between the gravitational wave energy density $\Omega_{\text{GW}}$ and the tensor power spectra $\mathcal{P}_\lambda$ as in Eq.~\eqref{eq:OmegaGW}.
In this work, ignoring scalar perturbations, we follow Refs.~\cite{Domenech:2021ztg,Fumagalli:2021mpc} and define them explicitly as
\begin{equation*}
    g_{ij} (t,\bm{x})=a(t)^2  \left[\delta_{ij}+h_{ij}(t,\bm{x})\right] \,,
\end{equation*}
\begin{equation*}
    \mathrm{e}_{ij}^{+}(\hat{\bm{k}})=\frac{\mathrm{e}_{i}(\hat{\bm{k}})\,\mathrm{e}_{j}(\hat{\bm{k}})-\bar{\mathrm{e}}_{i}(\hat{\bm{k}})\,\bar{\mathrm{e}}_{j}(\hat{\bm{k}})}{\sqrt{2}} \,,  \quad
    \mathrm{e}_{ij}^{\times}(\hat{\bm{k}})=\frac{\mathrm{e}_{i}(\hat{\bm{k}})\,\bar{\mathrm{e}}_{j}(\hat{\bm{k}})+\bar{\mathrm{e}}_{i}(\hat{\bm{k}})\,\mathrm{e}_{j}(\hat{\bm{k}})}{\sqrt{2}}
    \,,
\end{equation*}
where $\mathrm{e}_{i}(\hat{\bm{k}})$ are orthonormalized polarization vectors orthogonal to $\hat{\bm{k}}$. Note that these normalisations differ from Refs.~\cite{Inomata:2016rbd,Caprini:2018mtu}, hence the fact that numerical factors may differ.}
The tensor power spectrum is then defined by
\begin{equation}
    \braket{h^{\lambda}_{\bm{k}}(\eta)h^{\lambda'}_{\bm{k}'}(\eta)} = (2\pi)^3 \delta^{\lambda\lambda'} \delta(\bm{k} + \bm{k}') P_{\lambda}(\eta, k) \,.
\end{equation}
The observationally relevant quantity is the energy density per logarithmic wavelength of GWs, which can be determined from the power spectrum as
\begin{equation} \label{eq:OmegaGW}
\Omega_{\text{GW}}(\eta, k) = \frac{1}{12} \left( \frac{k}{a(\eta)H(\eta)}\right)^2 \sum_{\lambda=+,\times} \overline{\mathcal{P}_{\lambda}(\eta, k)} \,,
\end{equation}
where $\mathcal{P}_{\lambda} \equiv \frac{k^3}{2\pi^2}P_\lambda$ and the overline stands for an average over many oscillations of the GWs. This quantity is of course only meaningful for the tensor modes that have reentered the Hubble horizon, as it is only then that they behave as waves. The present-day GW density parameter is then given by $\Omega_{\text{GW},0}=c_g\Omega_{\text{r},0}\Omega_{\text{GW},\text{RD}}$ in terms of $\Omega_{\text{GW}}$ computed during radiation domination (RD) after the modes of interest have reentered the Hubble radius, the present radiation density parameter $\Omega_{\text{r},0}$, and a numerical factor $c_g\simeq 0.4$ that depends on the number of effective relativistic species at RD and today \cite{Caprini:2018mtu,Fumagalli:2021mpc}.

\subsection{Master integral}

At second order in perturbation theory, couplings of the schematic form $h\zeta\zeta$ in the Lagrangian lead to the following equation of motion for the tensor Fourier mode
\begin{equation}
\label{eq:EOM}
    h^{\lambda\prime\prime}_{\bm{k}}(\eta) + 2\mathcal{H} h^{\lambda\prime}_{\bm{k}}(\eta) + k^2h^\lambda_{\bm{k}}(\eta) = \mathcal{S}^\lambda_{\bm{k}}(\eta) \,,
\end{equation}
where $\mathcal{H}$ is the conformal-time Hubble parameter and the source term $\mathcal{S}^\lambda_{\bm{k}}(\eta)$ contains the second-order scalar terms. Concentrating on scalar-induced GWs, i.e.\ disregarding the homogeneous solution corresponding to the primordial contribution,
the solution of \eqref{eq:EOM} can be written as
\begin{equation}
    h^\lambda_{\bm{k}}(\eta) = \int_0^{\eta} \mathrm{d}\eta' G_{k} (\eta, \eta') \mathcal{S}_{\bm{k}}^{\lambda}(\eta') \,,
\end{equation}
with $G_{k}(\eta, \eta')$ being the equation's Green function, i.e.\ the solution of \eqref{eq:EOM} with an impulse source $\delta(\eta-\eta')$. One can then construct the tensor two-point function in order to obtain the following expression for the power spectrum:
\begin{equation}
\label{eq:GW_power_spectrum}
\begin{aligned}
P_{\lambda}(\eta,k)&=\frac{16}{k^4}\left(\frac{2}{3}\right)^4\int\frac{\dd^3 \bm{q}_1}{(2\pi)^3}\frac{\dd^3\bm{q}_2}{(2\pi)^3}\,Q_{\lambda}(\bm{k},\bm{q}_1)Q_{\lambda}(\bm{k},\bm{q}_2) \\
&\quad \times I\bigg(\frac{|\bm{k}-\bm{q}_1|}{k},\frac{q_1}{k},k\eta\bigg)I\bigg(\frac{|\bm{k}-\bm{q}_2|}{k},\frac{q_2}{k},k\eta\bigg) \braket{\zeta_{\bm{q}_1} \zeta_{\bm{k}-\bm{q}_1} \zeta_{-\bm{q}_2} \zeta_{-\bm{k} + \bm{q}_2}}' \,,
\end{aligned}
\end{equation}
in terms of the four-point function of the comoving primordial curvature perturbation $\zeta$, after expressing the Bardeen potential during radiation domination as $\Phi=\frac{2}{3}\,\zeta$, and as usual the notation $\langle\ldots\rangle'$ means that one has removed a factor of $(2\pi)^3$ and the delta function enforcing momentum conservation. In Eq.\ \eqref{eq:GW_power_spectrum} we introduced the function
\begin{equation}
    Q_\lambda(\bm{k}, \bm{q}) \equiv \mathrm{e}_{ij}^\lambda(\hat{\bm{k}})q_iq_j \,,
\end{equation}
whose properties will be discuss below, as well as the function
\begin{equation}
I(u,v,x)\equiv \int_{0}^{x}\dd x'\,kG_{k}(x/k,x'/k)f(u,v,x') \,,
\end{equation}
where $f(u,v,x)$ contains the information about the source. Explicit expressions for both $\mathcal{S}^\lambda_{\bm{k}}$ and $f$ can be found e.g.\ in \cite{Domenech:2021ztg}.\footnote{The functions $Q_\lambda$ and $I$ satisfy the following symmetry properties
\begin{equation*}
    Q_\lambda(\bm{k}, \bm{q}+c\bm{k})=Q_\lambda(\bm{k}, \bm{q}) \,,\qquad Q_\lambda(\bm{k}, \bm{q})=Q_\lambda(-\bm{k}, \bm{q})=Q_\lambda(\bm{k}, -\bm{q})\,,
\end{equation*}
\begin{equation*}
    I(u,v,x)=I(v,u,x) \,.
\end{equation*} Note that our function $I$ actually corresponds to the one of \cite{Kohri:2018awv,Adshead:2021hnm} multiplied by a factor $(2/3)^2$, or equivalently to the one used in \cite{Domenech:2021ztg}.}
We note that both $I$ and $f$ are dimensionless and defined with dimensionless arguments. To compute the time average in the formula \eqref{eq:OmegaGW} one needs the result  \nolinebreak\cite{Adshead:2021hnm}
\begin{equation} \label{eq:Isquared avg}
\begin{aligned}
\left(\frac{2}{3}\right)^4\lim_{x\to\infty}x^2\,\overline{I(u_1,v_1,x)I(u_2,v_2,x)}&=\frac{1}{2}\,I_A(u_1,v_1)I_A(u_2,v_2) \\
&\quad \times\Big[I_B(u_1,v_1)I_B(u_2,v_2)+I_C(u_1,v_1)I_C(u_2,v_2)\Big] \,,
\end{aligned}
\end{equation}
where
\begin{equation}
\begin{aligned}
I_A(u,v)&= \frac{3(u^2+v^2-3)}{4u^3v^3} \,,\\
I_B(u,v)&= -4uv+(u^2+v^2-3)\log\left|\frac{3-(u+v)^2}{3-(u-v)^2}\right| \,,\\
I_C(u,v)&= \pi(u^2+v^2-3)\Theta(u+v-\sqrt{3}) \,.
\label{IABC-functions}
\end{aligned}
\end{equation}
Once the four-point function of $\zeta$ is substituted in \eqref{eq:GW_power_spectrum}, the problem reduces to computing a six-dimensional integral. However, a simplification occurs when one considers the sum over the two helicity components of the tensor power spectrum. The following insert shows how, upon switching to spherical coordinates, the double integral over the azimuthal angles can be reduced to a single integral, regardless of the form of $\langle\zeta^4\rangle$.

\begin{framed}
{\small \noindent In this insert, we show that in the integral for $P_{\lambda}$, Eq.\ \eqref{eq:GW_power_spectrum}, one can trade the double azimuthal integration for a single integral. The four-point correlator is a function only of the magnitudes and scalar products of the momenta (as we will explain below). Working in a frame where $\bm{k} = k\hat{\bm{z}}$ and $\bm{q}_i = q_i\,(\sin\theta_i\cos\phi_i\,\hat{\bm{x}} + \sin\theta_i\cos\phi_i\,\hat{\bm{y}}+\cos\theta_i\,\hat{\bm{z}})$, one notes that the only dependence of the four-point function on the azimuthal angles $\phi_i$ comes from the scalar product $\bm{q}_1\cdot\bm{q}_2$, which involves $\cos(\phi_2 - \phi_1)$.

\vskip 4pt
Meanwhile, in this frame one also has
\begin{equation}
Q_{\lambda}(\bm{k},\bm{q}_i)=\frac{1}{\sqrt{2}}\,q_i^2\sin^2\theta_i\times\begin{cases}
\cos2\phi_i & \mbox{if $\lambda=+$} \\
\sin2\phi_i & \mbox{if $\lambda=\times$} \\
\end{cases}
\end{equation}
so that the sum over polarizations yields $\cos[2(\phi_2-\phi_1)]$ in the integrand. This produces a result of the form
\begin{equation}
    P_{h} \equiv \sum_{\lambda = +, \times}P_{\lambda} = \iint_0^{2\pi}\mathrm{d}\phi_1\mathrm{d}\phi_2 \cos[2(\phi_2-\phi_1)]\mathcal{F}\left[\cos(\phi_2-\phi_1)\right] \,,
\end{equation}
for some function $\mathcal{F}$. Doing the change of variables $\chi = \phi_1+\phi_2, \psi=\phi_1-\phi_2$, noting that the Jacobian is $1/2$ and being careful with the mapping of the integration domain, one eventually finds 
\begin{equation}
\label{eq: azimuthal dependence}
P_{h} = 2\pi \int_0^{2\pi} \mathrm{d}\psi\,\cos(2\psi) \mathcal{F}\left[\cos(\psi)\right] \,,
\end{equation}
and we will now refer to $\psi$ as \textit{the} azimuthal angle in the integral.
}
\end{framed}

\subsection{Scalar four-point function and integration kernels}
\label{sec:kernels}

The key ingredient in the master integral \eqref{eq:GW_power_spectrum} is the four-point function of the curvature perturbation, which carries information about the primordial physics. We decompose the correlator into disconnected and connected contributions,
\begin{equation}
    \braket{\zeta_{\bm{k}_1} \zeta_{\bm{k}_2} \zeta_{\bm{k}_3} \zeta_{\bm{k}_4}}' = \braket{\zeta_{\bm{k}_1} \zeta_{\bm{k}_2} \zeta_{\bm{k}_3} \zeta_{\bm{k}_4}}_{\text{d}}' + \braket{\zeta_{\bm{k}_1} \zeta_{\bm{k}_2} \zeta_{\bm{k}_3} \zeta_{\bm{k}_4}}_{\text{c}}' \,,
\end{equation}
and one recalls that the connected correlator vanishes when primordial fluctuations are drawn from a perfect Gaussian distribution.
Here we assume the presence of primordial non-Gaussianities as perturbative corrections to the Gaussian statistics, so that the connected piece of the four-point function---the primordial trispectrum---gives a non-zero contribution. Explicitly we have

\begin{equation}
\begin{aligned}
   \braket{\zeta_{\bm{q}_1} \zeta_{\bm{k}-\bm{q}_1} \zeta_{-\bm{q}_2} \zeta_{-\bm{k}+\bm{q}_2}}_{\text{d}}'&= (2\pi)^3\left[\delta(\bm{q}_1-\bm{q}_2)P_\zeta(q_1)P_\zeta(|\bm{k}-\bm{q}_1|)+\delta(\bm{q}_1+\bm{q}_2-\bm{k})P_\zeta(q_1)P_\zeta(q_2)\right] \,,\\
   \braket{\zeta_{\bm{q}_1} \zeta_{\bm{k}-\bm{q}_1} \zeta_{-\bm{q}_2} \zeta_{-\bm{k}+\bm{q}_2}}_{\text{c}}'&= T_{\zeta}(\bm{q}_1,\bm{k}-\bm{q}_1,-\bm{q}_2,-\bm{k}+\bm{q}_2) \,,
\end{aligned}
\end{equation}
where we disregarded in the first line the third Wick contraction $\propto \delta(\bm{k})$ that has no support on finite values of the external momentum $k$, and where one should understand the second line as a definition of the trispectrum.

\vskip 4pt
It is easy to see that the two terms in the disconnected correlator are equivalent upon integration. Thus we have
\begin{equation}
\label{eq: master formula vec(q) variables}
\begin{aligned}
    P_{\lambda, \text{d}} &= \frac{32}{k^4}\left(\frac{2}{3}\right)^4 \int \frac{\mathrm{d}^3 \bm{q}}{(2\pi)^3}\, Q_\lambda^2(\bm{k}, \bm{q})I^2\bigg(\frac{|\bm{k}-\bm{q}|}{k},\frac{q}{k},k\eta\bigg) P_\zeta(q)P_\zeta(|\bm{k} - \bm{q}|) \,,\\
    P_{\lambda, \text{c}} &= \frac{16}{k^4}\left(\frac{2}{3}\right)^4\int\frac{\dd^3 \bm{q}_1}{(2\pi)^3}\frac{\dd^3\bm{q}_2}{(2\pi)^3}\,Q_{\lambda}(\bm{k},\bm{q}_1)Q_{\lambda}(\bm{k},\bm{q}_2) \\
&\quad \times I\bigg(\frac{|\bm{k}-\bm{q}_1|}{k},\frac{q_1}{k},k\eta\bigg)I\bigg(\frac{|\bm{k}-\bm{q}_2|}{k},\frac{q_2}{k},k\eta\bigg)T_\zeta(\bm{q}_1,\bm{k}-\bm{q}_1,-\bm{q}_2,-\bm{k}+\bm{q}_2) \,.
\end{aligned}
\end{equation}
Changing variables as
\begin{equation}
    u_i=\frac{|\bm{k}-\bm{q}_i|}{k} \,,\qquad v_i=\frac{q_i}{k} \,,
\end{equation}
summing over polarizations and performing the time average with the help of formula \eqref{eq:Isquared avg}, we eventually obtain the disconnected and connected contributions to the dimensionless tensor power spectrum:
\begin{equation}
\label{eq: master formula uv variables}
\begin{aligned}
    \overline{\mathcal{P}_{h, \text{d}}} &=\left(\frac{aH}{k}\right)^2\int_0^{\infty}\dd v \int_{|1-v|}^{1+v} \dd u\; \mathcal{K}_{\text{d}}(u,v)\mathcal{P}_\zeta(ku)\mathcal{P}_\zeta(kv) \,,\\
    \overline{\mathcal{P}_{h, \text{c}}} &=\left(\frac{aH}{k}\right)^2\int_0^{\infty}\dd v_1 \int_{|1-v_1|}^{1+v_1} \dd u_1\int_0^{\infty}\dd v_2 \int_{|1-v_2|}^{1+v_2} \dd u_2\int_0^{2\pi}\dd\psi\; \mathcal{K}_{\text{c}}(u_1,v_1,u_2,v_2) \\
    &\quad\times \frac{\cos(2\psi)}{\pi}\mathcal{T}_\zeta(u_1,v_1,u_2,v_2,\psi) \,,\\
\end{aligned}
\end{equation}
where the dimensionless trispectrum function $\mathcal{T}_\zeta$
is defined as
\begin{equation}
    \mathcal{T}_\zeta(\bm{k}_1,\bm{k}_2,\bm{k}_3,\bm{k}_4)=\frac{(k_1k_2k_3k_4)^{9/4}}{(2\pi)^6}T_\zeta(\bm{k}_1,\bm{k}_2,\bm{k}_3,\bm{k}_4)\,,
\end{equation}
and where one has carried
out the above change of variables (and with $\psi=\phi_1-\phi_2$; see the insert above).
The integration kernels are given explicitly by
\begin{equation}
\label{eq:kernel}
\begin{aligned}
    \mathcal{K}_{\text{d}}(u,v)&=\left(\frac{4v^2-(1+v^2-u^2)^2}{2uv}\right)^2I_A^2(u,v)\left[I_B^2(u,v)+I_C^2(u,v)\right] \,,\\
    \mathcal{K}_{\text{c}}(u_1,v_1,u_2,v_2)&=\frac{1}{4(u_1v_1u_2v_2)^{5/4}}[4v_1^2-(1+v_1^2-u_1^2)^2][4v_2^2-(1+v_2^2-u_2^2)^2] \\
    &\quad\times I_A(u_1,v_1)I_A(u_2,v_2)\left[I_B(u_1,v_1)I_B(u_2,v_2)+I_C(u_1,v_1)I_C(u_2,v_2)\right] \,,
\end{aligned}
\end{equation}
where the functions $I_{A,B,C}$ were defined in Eq.~\eqref{eq:Isquared avg}.
Note that $\mathcal{K}_{\text{c}}(u,v,u,v)=\mathcal{K}_{\text{d}}(u,v)/\sqrt{uv}$, and that the kernel for the disconnected contribution was first derived in Refs.~\cite{Espinosa:2018euj,Kohri:2018awv}.

\vskip 4pt
It is useful to keep in mind the physical processes described by the kernel $\mathcal{K}_{\mathrm{c}}$.
In particular, in the above formulas, $\boldsymbol{q}_1$ and $\boldsymbol{k}-\boldsymbol{q}_1$ (respectively $-\boldsymbol{q}_2$ and $\boldsymbol{-k}+\boldsymbol{q}_2$) are the wavevectors of two scalar perturbations $\zeta$ that interact to generate a GWs of wavevector $\boldsymbol{k}$ (respectively $-\boldsymbol{k}$). Let us also recall that a given $\zeta$ mode stays constant outside the Hubble radius, before it decays and oscillates from Hubble re-entry onwards. These are the oscillations that can generate a substantial production of GWs, when the frequency $k$ of the graviton mode matches the one of the source term in Eq.~\eqref{eq:EOM}. This resonance occurs at $u_i+v_i=\sqrt{3}$, hence the divergence of $I_B$ at this location in Eq.~\eqref{IABC-functions}. 
On the other hand, when one of the momentum of the $\zeta$ modes is much smaller than $k$, i.e. when any one of the $(u_i,v_i)$ variables is $\ll 1$, the GW generation is negligible, as the source term is then suppressed by the negligible gradient of the corresponding soft $\zeta$ mode.
Eventually, in the opposite limit of a gravitational wave of momentum $k$ generated from two scalar fluctuations of comparatively hard momenta $q_i, |\bm{q}_i-\bm{k}| \gg k$, i.e.\ $(u_i,v_i) \gg 1$, the generation of GWs is also suppressed, as around the relevant time of Hubble re-entry for the $k$ mode, the corresponding $\zeta$ modes have already decayed substantially since their Hubble re-entry. The suppression of the GW production in the latter two limits will be useful in the following to restrict our analysis to the relevant kinematical domain, see also App.~\ref{app:kernel} for more details.

\subsection{Dissecting the trispectrum contribution}
\label{subsec:Dissecting the trispectrum}

In this section, we use the properties of the trispectrum and the symmetries of the six-dimensional integrand in Eq.~\eqref{eq: master formula vec(q) variables} to simplify the expression of the trispectrum-induced gravitational waves spectrum.

\vskip 4pt
First, we note that, being defined as the connected 4-point function of indistinguishable operators $\zeta_{\bm{k}_a}$, the trispectrum should really be understood as a function of the set $\{\bm{k}_1,\bm{k}_2,\bm{k}_3,\bm{k}_4\}$, independently of any number of permutations of the four external wave-vectors in the list $(\bm{k}_1,\bm{k}_2,\bm{k}_3,\bm{k}_4)$.
In practice, any quantum or classical calculation of the trispectrum gives several contributions corresponding to these 24 permutations:
\begin{equation}
    T_\zeta\left[\{\bm{k}_1,\bm{k}_2,\bm{k}_3,\bm{k}_4\}\right]=\tilde{T}_\zeta\left[(\bm{k}_1,\bm{k}_2,\bm{k}_3,\bm{k}_4)\right] + \,\, \text{23 perm.} \,,
\end{equation}
where the individual contributions $\tilde{T}_\zeta$ are not in general invariant under permutations of their arguments.\footnote{As we shall see, some trispectrum shapes do verify more symmetries under the exchanges of external momenta, which reduces the number of explicit permutations $\tilde{T}_\zeta$ needed to be written down.}
We now describe a useful way to arrange those 24 permutations:
\begin{itemize}
    \item Pick two pairs of momenta without ordering them. There are three such possible choices:
    $\{\{\bm{k}_1,\bm{k}_2\},\{\bm{k}_3,\bm{k}_4\}\}$, $\{\{\bm{k}_1,\bm{k}_3\},\{\bm{k}_2,\bm{k}_4\}\}$ and $\{\{\bm{k}_1,\bm{k}_4\},\{\bm{k}_2,\bm{k}_3\}\}$.
    Microscopically, these choices correspond respectively to the famous $s$, $t$ and $u$ channels of $2 \rightarrow 2$ scattering processes in particle physics, with each of the two pairs corresponding to the initial or final state.
    We can define $s=|\bm{k}_1+\bm{k}_2|$, $t=|\bm{k}_1+\bm{k}_3|$ and $u=|\bm{k}_1+\bm{k}_4|$ as the momentum of the exchanged particle for each of these channels.
    Although in cosmology there is no distinction between the initial and final states (all operators $\zeta_{\bm{k}_a}$ in the four-point function are evaluated at the end of inflation), we shall see that it is still useful to consider separately the contributions from the equivalent $s$, $t$ and $u$ channels, especially for trispectra with scalar-exchange microscopical origin.
    \item For each of these channels, it is possible to order one pair with respect to the other one (two choices, like $\left(\{\bm{k}_1,\bm{k}_2\},\{\bm{k}_3,\bm{k}_4\}\right)$ and $\left(\{\bm{k}_3,\bm{k}_4\},\{\bm{k}_1,\bm{k}_2\}\right)$ for the $s$-channel), and then to order internally each of the two pairs (two choices for each pair, so four choices in total).
    In total, this makes up for 8 individual contributions (that preserve the momentum of the exchanged particle) for each of the 3 channels, hence we correctly describe the 24 permutations.
\end{itemize}
We therefore write explicitly the three channels as
\begin{equation}
\label{eq: def channels}
    T_\zeta = \left( T_s + T_t +T_u \right)  + \,\, \text{7 perm.}  \,, \,\,\text{with}  \begin{cases}
      T_s = \tilde{T}_\zeta\left[(\bm{k}_1,\bm{k}_2;\bm{k}_3,\bm{k}_4)\right] \\
      T_t = \tilde{T}_\zeta\left[(\bm{k}_1,\bm{k}_3;\bm{k}_2,\bm{k}_4)\right] \\
      T_u = \tilde{T}_\zeta\left[(\bm{k}_1,\bm{k}_4;\bm{k}_2,\bm{k}_3)\right] 
    \end{cases}
    \,,
\end{equation}
where the remaining $7$ permutations preserve the momentum of the exchanged particle, i.e.\ the channel. We have also seen that the kernel (and the integral measure) which multiplies the trispectrum in Eq.~\eqref{eq: master formula vec(q) variables} is invariant under the following exchanges:
\begin{align}
\label{eq: symmetries}
    \bm{k}-\bm{q}_1 \leftrightarrow \bm{q}_1  
    &\iff \bm{k}_1 \leftrightarrow \bm{k}_2 \,\,\text{in} \,\, T_\zeta \,, \\
    \bm{k}-\bm{q}_2 \leftrightarrow \bm{q}_2  
    &\iff \bm{k}_3 \leftrightarrow \bm{k}_4 \,\,\text{in} \,\, T_\zeta \,, \nonumber \\
    \bm{q}_1 \leftrightarrow -\bm{q}_2 \,\, \text{and}\,\, \bm{k} \leftrightarrow -\bm{k} 
    &\iff (\bm{k}_1,\bm{k}_2) \leftrightarrow  (\bm{k}_3,\bm{k}_4) \,\,\text{in} \,\, T_\zeta  \nonumber \,.
\end{align}
Those symmetries enable one to drastically reduce the total number of permutations in the trispectrum needed for the computation.
Let us see how these permutations affect the three different channels, $s$, $t$ and $u$.
First, the 8 contributions of the $s$-channel are equivalent upon applying the above symmetries of the kernel, therefore the contribution from the $s$-channel is simply 8 times the one of the seed $T_s$.
For the $t$-channel, half of the contributions are equivalent to the seed $T_t$, while the other half does not preserve the channel upon the permutations in Eq.~\eqref{eq: symmetries} and can be shown to be equivalent to the other seed $T_u$.
Similarly for the $u$-channel, on the 8 contributions, 4 are equivalent to $T_u$ and 4 to $T_t$, once the symmetries are taken into account.
The trispectrum-induced gravitational waves spectrum can thus be written in terms of only three fundamental contributions corresponding to the seeds $T_s$, $T_t$ and $T_u$:
\begin{equation}
\label{eq: seeds contributions}
    \mathcal{P}_h^\mathrm{c}=8 \mathcal{P}_h^{\mathrm{c},s} + 4\left(\mathcal{P}_h^{\mathrm{c},t} + \mathcal{P}_h^{\mathrm{c},u} \right)
     + 4\left(\mathcal{P}_h^{\mathrm{c},t} + \mathcal{P}_h^{\mathrm{c},u} \right)
     =
     8\left( \mathcal{P}_h^{\mathrm{c},s} + \mathcal{P}_h^{\mathrm{c},t} + \mathcal{P}_h^{\mathrm{c},u}\right) \,,
\end{equation}
with $\mathcal{P}_h^{\mathrm{c},s}$ simply corresponding to $\mathcal{P}_h^{\mathrm{c}}$ with $T_\zeta$ replaced by $T_s$, etc.

\vskip 4pt
Finally, in conventional situations where parity is not violated, the trispectrum is only a function of the magnitudes of the four external momenta $k_a$ as well as the ones of the exchanged particles: $s$, $t$ and $u$.
This justifies \textit{a posteriori} the argument in the insert above, where we showed that the trispectrum-induced tensor power spectrum can be written as in Eq.~\eqref{eq: azimuthal dependence}, that the only dependence on the angles $\phi_{1,2}$ comes from the scalar product $\bm{q}_1 \cdot \bm{q}_2$.
Given the ordering of the arguments of the trispectrum shape in the master formula, Eq.~\eqref{eq: master formula vec(q) variables}, it is clear that this scalar product only appears in the variables $t$ and $u$ (and not $k_a$ nor $s$).
An important consequence of this fact is that any trispectrum shape that is not a function of $t$ or $u$ does not depend on the angle $\psi$ and therefore gives a vanishing contribution to the tensor power spectrum, by virtue of Eq.~\eqref{eq: master formula uv variables}: $\int_{0}^{2\pi} \mathrm{d}\psi \, \mathcal{K}_c(u_1,v_1,u_2,v_2)  \mathrm{cos}(2\psi) \mathcal{T}_\zeta(u_1,v_1,u_2,u_2,$\sout{$\psi$}$) =0$.

\subsection{Trispectrum shapes}

So far in the literature, only the effects of \textit{local} non-Gaussianities have been taken into account for the generation of scalar-induced gravitational waves, for inflationary models with strong deviations from scale-invariance.
In this work, we thoroughly investigate the effect of the trispectrum in more conventional situations where scale invariance is observed.
Here, we explain the physical origin and properties of the trispectrum shapes that we shall consider as potential sources for the gravitational wave energy density.

\paragraph{Regular shapes: derivative interactions in single-field inflation.}
A general class of effectively single-field inflationary models is described by the Effective Field Theory of inflationary fluctuations~\cite{Creminelli:2006xe, Cheung:2007st}.
In this context and assuming a slow-varying background, the scalar trispectrum has been computed taking into account both the contact quartic interactions of $\zeta$ and the scalar-exchange processes corresponding to two cubic interactions with one exchanged $\zeta$-mode~\cite{Chen:2009bc}.
These interactions correspond to derivative interactions and lead to regular shapes, i.e.\ without singularities in particular kinematic configurations.
Schematically, the resulting trispectrum is (see Ref.~\cite{Chen:2009bc} for the explicit formulas)
\begin{align}
\label{eq: regular shape}
   T_\zeta^\mathrm{reg} &= T^\mathrm{contact}(k_a,s,t,u)+T^\mathrm{exchange}(k_a,s,t,u)\,,\quad \text{with}  \\ T^\mathrm{contact}(k_a,s,t,u) &= T_{\zeta^{\prime 4}}(k_a) +  T_{\zeta^{\prime 2} (\partial\zeta)^2}(k_a,s,t,u) + T_{(\partial\zeta)^4}(k_a,s,t,u)  \nonumber  \,,\quad \text{and}\\ T^\mathrm{exchange}(k_a,s,t,u) &=
    T_{\zeta^{\prime3} - \zeta^{\prime3}}(k_a,s,t,u) + T_{\zeta^{\prime 3} - \zeta^{\prime} (\partial\zeta)^2}(k_a,s,t,u) + T_{ \zeta^{\prime} (\partial\zeta)^2 - \zeta^{\prime} (\partial\zeta)^2}(k_a,s,t,u) \nonumber \,,
\end{align}
where we have dissected the contributions from different interactions. From these expressions, and given the remark in the previous section that trispectrum shapes that do not depend on $t$ nor $u$ give a vanishing contribution to the trispectrum-induced tensor power spectrum for lack of azimuthal dependence in the integral, we can already note that the quartic interaction $\zeta^{\prime 4}$ does not induce gravitational waves at all.
Moreover, some sub-contributions (not displayed explicitly here) of the remaining 5 terms depend only on $s$, and not $t$ nor $u$, and therefore do not contribute either.

\paragraph{Local shape: massless scalar fields.}
It is possible that the dominant contribution to the primordial trispectrum comes from non-derivative interactions of classical, super-horizon fields, in which case non-Gaussianities are said to be of the \textit{local} type (see Ref.~\cite{Wands:2010af} for a review).
If the primordial matter content is made of massless scalar fields, it is possible to distinguish two contributions in the resulting trispectrum:
\begin{align}
\label{eq: local shape}
    T_\zeta^\mathrm{loc} &= \tau_\mathrm{NL} \left[\left(P_\zeta(s) P_\zeta(k_{1}) P_\zeta(k_{3}) 
    + \,\, \text{3 perm.} \right)  + \,\,  (s \leftrightarrow t)  +  (s \leftrightarrow u)  \right] \nonumber \\
    & \quad + \frac{54}{25} g_\mathrm{NL} \left[P_\zeta(k_{1}) P_\zeta(k_{2}) P_\zeta(k_{3}) + \,\, \text{3 perm.}  \right] \,,
\end{align}
where $\tau_\mathrm{NL}$ and $g_\mathrm{NL}$ are two dimensionless parameters measuring the size of the trispectrum.

\vskip 4pt
The $\tau_\mathrm{NL}$-contribution corresponds microscopically to the exchange of scalar particles through two cubic interactions, and the implicit permutations in Eq.~\eqref{eq: local shape} concern only the internal ordering of the pairs within a channel, e.g. $\{(\bm{k}_1,\bm{k}_2),(\bm{k}_3,\bm{k}_4)\}$ for the first term corresponding to the $s$-channel. In the particular case of single-field inflation, the scalar-exchange trispectrum amplitude is set by the size of the bispectrum, $\tau_\mathrm{NL}=\left(6 f_\mathrm{NL}/5\right)^2$, while more generally it should satisfy the so-called Suyama-Yamaguchi inequality~\cite{Suyama:2007bg}: $\tau_\mathrm{NL} \geqslant \left(6 f_\mathrm{NL}/5\right)^2$. The scalar-exchange contribution diverges not only for infinitely soft external momenta, but also for so-called collapsed configurations corresponding to infinitely soft exchanged momentum: $s=|\bm{k}_1+\bm{k}_2| \rightarrow 0$ for the $s$-channel, etc.
Note also that the $s$, $t$ and $u$ channels depend respectively only on $s$, $t$ and $u$ (and all depend on all $k_a$).
Therefore the $s$-channel of the scalar-exchange local shape cannot contribute to the tensor power spectrum.

\vskip 4pt
The $g_\mathrm{NL}$-term corresponds microscopically to a contact quartic interaction and does not make the difference between the $s$, $t$ and $u$ channels.
It is independent from the size of the bispectrum, and diverges only for soft external momenta.
It does not depend on $t$ nor $u$, and therefore has no azimuthal dependence. Thus, it does not contribute at all to the secondary generation of gravitational waves.

\paragraph{Extended local shape: light and spinning fields.}
It has been shown that the trispectrum corresponding to the exchange of massive (but light) and spinning (but bosonic) fields could be described by the following ansatz, in kinematical configurations close enough to the respective collapsed limits of the different channels~\cite{Assassi:2012zq,Bordin:2019tyb}:
\begin{equation}
\label{eq: extended local shape}
    T^{\mathrm{loc},\Delta, S}_{\zeta} \simeq  \begin{cases}
      \tau_{\text{NL}} \left[\left( \frac{s^2}{k_1k_3}\right)^{\Delta} P_\zeta(s)P_\zeta(k_1)P_\zeta(k_3) \mathrm{P}_{S}\left(\hat{\bm{k}}_1 \cdot\hat{\bm{k}}_3\right) + \,\, \text{3 perm.} \right] \text{for $s \rightarrow 0$} 
      \\ \\
     \tau_{\text{NL}} \left[\left( \frac{t^2}{k_1k_2}\right)^{\Delta} P_\zeta(t)P_\zeta(k_1)P_\zeta(k_2) \mathrm{P}_{S}\left(\hat{\bm{k}}_1 \cdot\hat{\bm{k}}_2\right) + \,\, \text{3 perm.} \right] \text{for $t \rightarrow 0$}
     \\ \\
     \tau_{\text{NL}} \left[\left( \frac{u^2}{k_1k_3}\right)^{\Delta} P_\zeta(u)P_\zeta(k_1)P_\zeta(k_3) \mathrm{P}_{S}\left(\hat{\bm{k}}_1 \cdot\hat{\bm{k}}_3\right) + \,\, \text{3 perm.} \right] \text{for $u \rightarrow 0$}\,,
    \end{cases}
\end{equation}
where $\Delta =3/2 - \sqrt{9/4 - (m/H)^2}$ is the dimensionless mass parameter of the new particle with mass $m$, $S$ is its spin and $\mathrm{P}_S$ is the corresponding Legendre polynomial of order $S$.
We have also defined the unit vectors as $\hat{\bm{k}}=\bm{k}/k$, i.e.\ the Legendre polynomial is only a function of the angle between the two momenta appearing in its argument.
This extended local shape still diverges in the collapsed limit for light enough exchanged fields, and gets a different angle-dependence due to the Legendre polynomials when those fields are spinning.
For example, $\hat{\bm{k}}_1 \cdot\hat{\bm{k}}_3 = (t^2- k_1^2-k_3^2)/(2k_1k_3)$, and therefore the $s$-channel depends explicitly on the variable $t$ (and $u$ for other implicit permutations in the first line of Eq.~\eqref{eq: extended local shape}) if the exchanged particle is spinning.
Therefore in this case, even the $s$-channel should \textit{a priori} contribute to the gravitational wave energy density.

\vskip 4pt
Let us comment about a possible extension of the ``$g_\mathrm{NL}$ shape" of local non-Gaussianities to the case of the exchange of massive and spinning fields. In the case of massive but spinless fields, the trispectrum has been computed in such soft limit for external momenta $k_a \rightarrow 0$ (rather than internal momenta $s,t,u \rightarrow 0$ for the collapsed limit)~\cite{Chen:2018sce}. Since no new angular dependence arises, it cannot contribute to the gravitational wave energy density.
As for spinning fields, for which one could expect a more complex angular dependence, we found no explicit computation of the trispectrum in this kinematical limit.
However, it can be argued that its contribution to the gravitational wave energy density should anyway be negligible.
Indeed, such trispectrum shapes are the largest in the limit of soft external momenta, corresponding in the integrals of Eq.~\eqref{eq: master formula uv variables} to either of the variables $u_i, v_i \rightarrow 0$. But, as explained at the end of Section \ref{sec:kernels}, the integration kernel $\mathcal{K}_\mathrm{c}$ is damped in that limit. This corresponds to the negligible sourcing of GWs of momentum $k$ by the interactions of a $\zeta$ mode of the same wavelength and of a comparatively much larger wavelength $\zeta$ mode, whose gradients are suppressed. As a result, one can infer that the resulting contribution should be negligible, independently of the precise form of the shape.

\section{Phenomenology of a large scalar trispectrum}
\label{sec:pheno}

We now consider the effect of the various trispectrum shapes defined in the previous section on the secondary generation of gravitational waves.
We focus here on the simplest possibility of a scale-invariant theory, i.e.\ with negligible time-dependence of the background leading to trispectra of the type of Eq.~\eqref{eq: regular shape} for derivative cubic and quartic interactions, or Eqs.~\eqref{eq: local shape}--\eqref{eq: extended local shape} with $P_\zeta(k) = (2\pi^2/k^3) \mathcal{P}_\zeta$ and constant $\mathcal{P}_\zeta$ in the case of local non-Gaussianities.
For these scale-invariant theories, we first quote the disconnected (Gaussian) contribution to the gravitational wave energy density during the radiation era. It reads
\begin{equation}
    \Omega_{\mathrm{GW},\mathrm{d}} = \frac{1}{12} \left(\frac{k}{a H}\right)^2 \overline{\mathcal{P}_{h, \mathrm{d}}}  =  \mathcal{P}_\zeta^2 \mathcal{I}_{\text{d}} \quad \text{with} \quad \mathcal{I}_{\mathrm{d}} = 3.20\,,
\end{equation}
where $\mathcal{I}_\mathrm{d}$ is simply the numerical result of the dimensionless integral over the variables $(v,u)$ in the disconnected contribution, as shown in Eq.~\eqref{eq: master formula uv variables}. 
Of course, the scalar power spectrum being scale-invariant, so is the secondary sourced gravitational wave energy density.

\subsection{Equilateral shapes from the EFT of inflation}

In inflationary scenarios with an effectively single-field theory for fluctuations, the trispectrum may become non-negligible due to large derivative cubic and quartic interactions.
In the following, we use the notations $c_s^2, \lambda, \Sigma, \mu$ for the cubic and quartic couplings of $P(X)$ models of inflation~\cite{Garriga:1999vw}, where $X=-\frac12 g^{\mu\nu} \partial_\mu \phi \partial_\nu \phi$ is the kinetic term of the inflaton $\phi$, because it is in this context that the trispectrum shape has been fully computed.
We stress however that our results readily apply to the largest class of single-field effective field theories (EFTs) for fluctuations, where the cubic and quartic couplings are parameterized by the numbers $c_s^2, A, B$, independently of whether they are considered from a bottom-up~\cite{Creminelli:2006xe, Cheung:2007st} or a top-down (see, e.g.~\cite{Tolley:2009fg,Cremonini:2010ua,Achucarro:2012sm,Garcia-Saenz:2019njm,Pinol:2020kvw}) perspective, with the following replacement rule\footnote{In this parametrization of the Wilson coefficients of the EFT, the parameter $A$ multiplies the cubic interaction $\zeta^{\prime3}$ (and higher-point terms) and the parameter $B$ multiplies the quartic interaction $\zeta^{\prime4}$ (and higher-point terms). All other interactions, namely $\zeta'(\partial\zeta)^2$ at cubic order and $\zeta^{\prime2}(\partial\zeta)^2,(\partial\zeta)^4$ at quartic order, are ``universal'' in the sense that they are present even when $A$ and $B$ are tuned to zero, see Eq.~3.82 in \cite{Pinol:2021nha}.}
\begin{align} \label{eq:wilson coeffs}
    2\frac{\lambda}{\Sigma}&\equiv \frac{2X^2P_{,XX}+4/3 X^3P_{,XXX}}{XP_{,X}+2X^2P_{,XX}} \Longleftrightarrow -\left(\frac{1}{c_s^2}-1\right) A \,, \\
    \frac{\mu}{\Sigma} &\equiv \frac{1/2 X^2P_{,XX}+2X^3P_{,XXX}+2/3X^4P_{,XXXX}}{XP_{,X}+2X^2P_{,XX}} \Longleftrightarrow \left(\frac{1}{c_s^2}-1\right) \frac{B}{c_s^2} \,. \nonumber
\end{align}
In these contexts, assuming a slow-varying background, one can derive the trispectrum shape $T_\zeta^\mathrm{reg}$ of Eq.~\eqref{eq: regular shape} (that, we recall, can be found explicitly in Ref.~\cite{Chen:2009bc}), which once plugged in Eq.~\eqref{eq: master formula uv variables} gives
\begin{align}
\label{eq: result regular shape}
    \Omega_{\mathrm{GW},\mathrm{c}} &= \frac{1}{12} \left(\frac{k}{a H}\right)^2 \overline{\mathcal{P}_{h, \mathrm{c}}} 
     = \mathcal{P}_\zeta^3 \left[\left( \frac{1}{c_s^2}-1 \right)   \mathcal{I}_{c 3} 
    + \left(\frac{\lambda}{\Sigma}\right)^2 \mathcal{I}_{s1}
    + \frac{\lambda}{\Sigma}\left( \frac{1}{c_s^2}-1 \right)   \mathcal{I}_{s2}    
    + \left( \frac{1}{c_s^2}-1 \right)^2  \mathcal{I}_{s3} 
    \right]\,,  \nonumber \\
    \text{with} & 
    \,\, \mathcal{I}_{c3} = 0.198\,,
    \,\, \mathcal{I}_{s1} = -0.015\,,
    \,\, \mathcal{I}_{s2} = 0.217\,,
    \,\, \mathcal{I}_{s3} = 0.515\,,
\end{align}
where we have computed numerically the dimensionless integrals over the five variables $(v_1, u_2 , v_2,$ $u_2, \psi)$ to find the numbers $\mathcal{I}$, see App.~\ref{app:numerics} for more details about the numerics.
Note that, as anticipated, the contact interaction $\zeta^{\prime4}$ (the would-be ``$c1$" term) does not contribute for lack of $t, u$ dependence.
Moreover, the contact interaction $\zeta^{\prime2}(\partial \zeta)^2$ (the would-be ``$c2$" term) also does not contribute due to its simple azimuthal dependence on the variable $\psi$; the integral over $\psi$ averages to 0 when this particular trispectrum shape is multiplied by $\mathrm{cos}(2\psi)$ in Eq.~\eqref{eq: master formula uv variables}.
The other contact interaction and the three scalar-exchange diagrams all contribute.
Note also that the ratio between each contribution $(i) \in \{c3, s1, s2, s3\}$ and the gaussian-induced gravitational wave energy density may be written as
\begin{tcolorbox}[colframe=white,arc=0pt,colback=greyish2]
\begin{equation}
\label{eq: ratio regular shape}
    \frac{\Omega_{\mathrm{GW},\mathrm{c}}}{\Omega_{\mathrm{GW},\mathrm{d}}} = \sum_{(i)}  \frac{\mathcal{I}_{(i)}}{\mathcal{I}_\mathrm{d}} 
    t_\mathrm{NL}^{(i)} \mathcal{P}_\zeta
    \,,
\end{equation}
\end{tcolorbox}
\noindent with $t_\mathrm{NL}^{(i)} \in \left\{1/c_s^2-1, (\lambda / \Sigma)^2, \lambda/\Sigma \times (1/c_s^2-1), (1/c_s^2-1)^2 \right\}$ the dimensionless parameters appearing in Eq.~\eqref{eq: result regular shape}, that can be related to the typical size of the scalar trispectrum for the four contributions $c3, s1, s2, s3$, evaluated in regular configurations of the tetrahedron formed by the four external wavevectors: $\forall a \,, \,\, k_a=s=u$.
Because the ratios $\mathcal{I}_{(i)}/\mathcal{I}_\mathrm{d}$ are smaller than unity, one concludes that the size of the trispectrum in regular configurations, $t_\mathrm{NL}^{(i)}$, directly determines the relative amplitude of the connected to the disconnected contributions to the gravitational wave energy density.
Therefore, as should be clear from this expression, only a very large trispectrum, $t_\mathrm{NL}^{(i)} \sim \mathcal{P}_\zeta^{-1}$, can induce a spectrum of gravitational waves comparable to the one from the Gaussian part of scalar fluctuations.
Actually, note that although $\lambda/\Sigma$ is a priori independent of $1/c_s^2$, not all sizes of this cubic coupling are radiatively stable. In this respect, the notations of the EFT of inflation are more informative, precisely because the dimensionless numbers $A, B$ have been defined such that their natural values are of order unity for a low speed of sound $c_s$.
For example, cubic interactions proportional to $1/c_s^2$ induce a loop correction to $A$ of order unity, hence values of $A$ close to unity are radiatively stable.
These considerations have two implications:
\begin{itemize}
    \item The scalar-exchange scalar trispectra always induce larger contributions to the gravitational wave energy density than the contact scalar trispectra (that either vanish for $c1$ and $c2$, or is suppressed by $c_s^2$ for natural values of the coupling constants, for $c3$);
    \item The only way to get a contribution from a scalar trispectrum due to derivative interactions, that is larger than from the disconnected Gaussian scalar contribution, is to have a very small speed of sound, $1/c_s^4 \gg \mathcal{P}_\zeta^{-1}$.
\end{itemize}
However, as we will see in Sec.~\ref{sec: perturbativity bound}, such large values of the effective coupling $1/c_s^2$ are precisely the ones that violate unitarity bounds and therefore do not allow for the usual perturbative treatment of fluctuations that has been used throughout the computation of Eq.~\eqref{eq: ratio regular shape}.

\subsection{Local non-Gaussianities}

The local trispectrum may be large if sufficiently light (effectively massless) scalar fields have non-trivial interactions on super-horizon scales.
In that case, both the $\tau_\mathrm{NL}$ and the $g_\mathrm{NL}$ contributions in $T_\zeta^\mathrm{loc}$ of Eq.~\eqref{eq: local shape} may be large.
However, we have already commented that the contact quartic interactions encoded in $g_{\text{NL}}$ does not contribute to the gravitational wave energy density, for the lack of azimuthal dependence of the corresponding trispectrum shape.
Therefore, only the scalar-exchange $\tau_\mathrm{NL}$ term may contribute.
We also recall that anyway the entire $s$-channel gives a vanishing contribution.
We can thus focus on the contributions from the $t$ and $u$ channels of the $\tau_\mathrm{NL}$ term.
Interestingly, due to the presence of divergences in physically relevant kinematical configurations in those trispectrum shapes, when respectively $t$ and $u$ become small compared to $k$, the computation of the corresponding signal can be carried analytically under some approximation, a task to which we now turn.

\vskip 4pt
First, the seeds for the $t$ and $u$ channels of the $\tau_\mathrm{NL}$ term in $T_\zeta^\mathrm{loc}$ of Eq.~\eqref{eq: local shape}, are
\begin{align}
    T_t = \frac{\tau_\mathrm{NL}}{2} P_\zeta(t) P_\zeta(k_{1}) P_\zeta(k_{2}) \,, \quad
    T_u = \frac{\tau_\mathrm{NL}}{2} P_\zeta(u) P_\zeta(k_{1}) P_\zeta(k_{3}) \,.
\end{align}
Plugging these expression into the master formula Eq.~\eqref{eq: master formula vec(q) variables}, changing variable in the $\bm{q}_2$ integral into respectively $\bm{t}=\bm{q}_1-\bm{q}_2$ and $\bm{u}=\bm{q}_1+\bm{q}_2-\bm{k}$, we get the following contributions from the seeds
\begin{align}
\label{Pct}
P_{h}^{\mathrm{c},t} = \frac{16}{k^4} \left(\frac{2}{3}\right)^4 \int \frac{\mathrm{d}^3\bm{q}}{(2\pi)^3}\int \frac{\mathrm{d}^3\bm{t}}{(2\pi)^3} &\sum_{\lambda=+,\times} Q_\lambda(\bm{k}, \bm{q})Q_\lambda(\bm{k}, \bm{q}-\bm{t})\\
&I\left(\frac{|\bm{k} - \bm{q}|}{k}, \frac{q}{k}, k\eta \right)I\left(\frac{|\bm{k} - (\bm{q}-\bm{t})|}{k}, \frac{|\bm{q}-\bm{t}|}{k}, k\eta \right) \nonumber \\
&\times \,\frac{\tau_{\text{NL}}}{2} P_\zeta(t)P_\zeta(q)P_\zeta(|\bm{k} - \bm{q}|) \,, \nonumber \\
P_{h}^{\mathrm{c},u} = \frac{16}{k^4} \left(\frac{2}{3}\right)^4 \int \frac{\mathrm{d}^3\bm{q}}{(2\pi)^3}\int \frac{\mathrm{d}^3\bm{u}}{(2\pi)^3} &\sum_{\lambda=+,\times} Q_\lambda(\bm{k}, \bm{q})Q_\lambda(\bm{k}, \bm{q} + \bm{u})\\
&I\left(\frac{|\bm{k} - \bm{q}|}{k}, \frac{q}{k}, k\eta \right)I\left(\frac{|\bm{k} - (\bm{q}+\bm{u})|}{k}, \frac{|\bm{q}+\bm{u}|}{k}, k\eta \right)\nonumber \\
&\times \, \frac{\tau_{\text{NL}}}{2} P_\zeta(u)P_\zeta(q)P_\zeta(|\bm{k} - (\bm{q}+\bm{u})|) \,.\nonumber 
\end{align}
At first sight, there seems to be three kinematical regions that could contribute substantially to the tensor power spectrum, namely when either one of the wavenumbers in the product of the three scalar power spectra $P_\zeta(p) \propto 1/p^3$ is soft (compared to $k$). However, considering the $t$ channel for definiteness, remember from the discussion in Section \ref{sec:kernels} that the GW production is suppressed for $q$ and $|\bm{k} - \bm{q}| \ll k$ (corresponding to small arguments in the first $I$ function in Eq.~\eqref{Pct}).
Hence, one can identify that the corresponding integral is dominated by only one kinematical limit, corresponding to the exchanged momentum $t$ (and $u$ for the other channel) being soft compared to $k$, something  that we will confirm numerically in the next section.
We therefore probe the behaviour of these integrals around the collapsed limit $t, u\rightarrow 0$.
First, for a fixed $\bm{q}$, we note that by sending $t, u\rightarrow 0$ the kernel becomes independent of these exchanged momenta.
For example, the contribution for the $t$-channel seed in this limit is
\begin{align}
P_{h}^{\mathrm{c},t} \simeq \frac{16}{k^4} \left(\frac{2}{3}\right)^4\int \frac{\mathrm{d}^3 \bm{q}}{(2\pi)^3} &\sum_{\lambda=+,\times}
Q_\lambda^2(\bm{k}, \bm{q}) I^2\left(\frac{|\bm{k} - \bm{q}|}{k}, \frac{q}{k}, k\eta \right)P_\zeta(q)P_\zeta(|\bm{k} - \bm{q})|)
 \\
&\times \, \frac{\tau_{\text{NL}}}{2} \int \frac{\mathrm{d}^3 \bm{t}}{(2\pi)^3} P_\zeta(t) \,, \nonumber
\end{align}
where one clearly recognizes in the first line, the disconnected contribution from Gaussian scalar perturbations to the tensor power spectrum, $P_h^\mathrm{d}$, up to a factor of $1/2$, see the master formula, Eq.~\eqref{eq: master formula vec(q) variables}.
Therefore, one finds $P_{h}^{\mathrm{c},t} \simeq (\tau_\mathrm{NL}/4) P_h^\mathrm{d} \int [\dd^3 \bm{t} / (2\pi)^3]  P_\zeta(t)$, and using the expression of the total trispectrum-induced tensor power spectrum in terms of the seeds contributions, see Eq.~\eqref{eq: seeds contributions}, one finds
\begin{equation}
P_{h}^{\mathrm{c}} = 8 \left(P_{h}^{\mathrm{c},t} +P_{h}^{\mathrm{c},u}\right) \simeq 2\tau_{\text{NL}} \,P_{h}^{\mathrm{d}} \left[ \int \frac{\mathrm{d}^3 \bm{t}}{(2\pi)^3}\, P_\zeta (t) + \int \frac{\mathrm{d}^3 \bm{u}}{(2\pi)^3}\, P_\zeta (u) \right] \,.
\end{equation}
Taking a scale-invariant power spectrum, $P_\zeta(k) = (2\pi^2/k^3)\mathcal{P}_\zeta$, one can note that the integral is formally log-divergent. Naturally, it is not physical to consider infinitely large wavelength fluctuations, as modes with wavelengths larger than our observable universe should not be described as fluctuations but rather as part of the background. Hence considering an IR cutoff scale $L^{-1}$ on allowed comoving wavenumbers, one eventually finds
\begin{tcolorbox}[colframe=white,arc=0pt,colback=greyish2]
\begin{equation}
     \frac{\Omega_{\mathrm{GW},\mathrm{c}}}{\Omega_{\mathrm{GW},\mathrm{d}}}  \simeq 4  \tau_{\text{NL}} \mathcal{P}_{\zeta} \log{kL} \,.
\label{eq:ratio}     
\end{equation}
\end{tcolorbox}
\noindent Again, the only way to make this contribution large, is to have large primordial non-Gaussianities, $\tau_\mathrm{NL} \log kL \gg \mathcal{P}_\zeta^{-1}$.
Although one may think that the large logarithm may help to make the signal larger at small scales $k L \gg 1$, we shall see that the perturbativity bound also includes it and constrains the connected contribution to be smaller than the disconnected one.

\subsection{Non-Gaussianities from massive and spinning particles}
\label{subsec: extended local shapes}

We now turn to the computation of the contribution from the extended local shape of trispectrum, due to the exchange of massive and/or spinning particles.
Intuitively, we should expect the contribution from heavier particles to be subdominant compared to the massless case, which we confirm in this section.
More precisely, we show the analytical dependence on the mass of this exchanged particle, and confirm this result numerically.
Eventually, we consider the exchange of spinning bosonic particles and show that although the relative contributions from the different channels are reshuffled, the overall contribution is always smaller than the one from the exchange of a massless scalar.

\vskip 4pt
For convenience, we remind the seeds for the $s$, $t$ and $u$ channels to which this extended local shape reduces close to their respective collapsed configurations, see Eq.~\eqref{eq: extended local shape}:
\begin{align}
    T_{s} & = \frac{\tau_{\text{NL}}}{2}\, \left( \frac{s^2}{k_1k_3}\right)^{\Delta} P_\zeta(u)P_\zeta(k_1)P_\zeta(k_3) \mathrm{P}_{S}(\hat{\bm{k}}_1\cdot\hat{\bm{k}}_3)\,, \nonumber \\
    T_{t} & = \frac{\tau_{\text{NL}}}{2}\, \left( \frac{t^2}{k_1k_2}\right)^{\Delta} P_\zeta(t)P_\zeta(k_1)P_\zeta(k_2) \mathrm{P}_{S}(\hat{\bm{k}}_1\cdot\hat{\bm{k}}_2)\,, \nonumber  \\ 
    T_{u} & = \frac{\tau_{\text{NL}}}{2}\, \left( \frac{u^2}{k_1k_3}\right)^{\Delta} P_\zeta(u)P_\zeta(k_1)P_\zeta(k_3) \mathrm{P}_{S}(\hat{\bm{k}}_1\cdot\hat{\bm{k}}_3)\,, \nonumber
\end{align}
where $\Delta = 3/2 - \sqrt{9/4 - (m/H)^2}$ and $\mathrm{P}_S$ is the Legendre polynomial of order $S$.

\paragraph{Spin-0 massive case.}
Let us first consider the spin-$0$ massive exchanged particle case, for which the Legendre polynomial is simply one, and therefore the $s$-channel seed gives a vanishing contribution again, for lack of azimuthal dependence.
Similarly to the massless case, switching to $t$ and $u$ variables, the $t$- and $u$-channels seeds contributions become
\begin{align}
P_{h}^{\mathrm{c},t} = \frac{16}{k^4} \left(\frac{2}{3}\right)^4 \int \frac{\mathrm{d}^3 \bm{q}}{(2\pi)^3}\int \frac{\mathrm{d}^3 \bm{t}}{(2\pi)^3} &\sum_{\lambda=+,\times} Q_\lambda(\bm{k}, \bm{q})Q_\lambda(\bm{k}, \bm{q}-\bm{t})\\
&I\left(\frac{|\bm{k} - \bm{q}|}{k}, \frac{q}{k}, k\eta \right)I\left(\frac{|\bm{k} - (\bm{q}-\bm{t})|}{k}, \frac{|\bm{q}-\bm{t}|}{k}, k\eta \right) \nonumber \\
&\times \,\frac{\tau_{\text{NL}}}{2}
\left(\frac{t^2}{q|\bm{k}-\bm{q}|}\right)^\Delta
P_\zeta(t)P_\zeta(q)P_\zeta(|\bm{k} - \bm{q}|) \,, \nonumber \\
P_{h}^{\mathrm{c},u} = \frac{16}{k^4} \left(\frac{2}{3}\right)^4 \int \frac{\mathrm{d}^3\bm{q}}{(2\pi)^3}\int \frac{\mathrm{d}^3\bm{u}}{(2\pi)^3} &\sum_{\lambda=+,\times} Q_\lambda(\bm{k}, \bm{q})Q_\lambda(\bm{k}, \bm{q} + \bm{u})\\
&I\left(\frac{|\bm{k} - \bm{q}|}{k}, \frac{q}{k}, k\eta \right)I\left(\frac{|\bm{k} - (\bm{q}+\bm{u})|}{k}, \frac{|\bm{q}+\bm{u}|}{k}, k\eta \right)\nonumber \\
&\times \, \frac{\tau_{\text{NL}}}{2} \left(\frac{u^2}{q|\bm{k}-(\bm{q}+\bm{u})|}\right)^\Delta P_\zeta(u)P_\zeta(q)P_\zeta(|\bm{k} - (\bm{q}+\bm{u})|) \,.\nonumber 
\end{align}
Although the integrals are now convergent for any $\Delta \neq 0$, the integrand is still considerably larger in the collapsed configurations, respectively $t, u \rightarrow 0$, to which we should restrict anyway by consistency of the regime of validity of these trispectrum shapes.
We therefore introduce a cutoff for the integrals over these exchanged momenta
\begin{equation}
    t_\star = u_\star = \alpha \sqrt{q|\bm{k}-\bm{q}|}\,,
\end{equation}
with $\alpha$ a dimensionless variable.
The smaller $\alpha$, the more we restrict the domain of integration close to the collapsed configuration, so we consider $\alpha < 1$ by consistency.
Close to this collapsed limit, $\alpha \ll 1$, we find for example for the seed of the $t$-channel:
\begin{align}
P_{h}^{\mathrm{c},t} \simeq \frac{16}{k^4} \left(\frac{2}{3}\right)^4\int \frac{\mathrm{d}^3 \bm{q}}{(2\pi)^3} &\sum_{\lambda=+,\times}
Q_\lambda^2(\bm{k}, \bm{q}) I^2\left(\frac{|\bm{k} - \bm{q}|}{k}, \frac{q}{k}, k\eta \right)P_\zeta(q)P_\zeta(|\bm{k} - \bm{q})|)
 \\
&\times \, \frac{\tau_{\text{NL}}}{2} \int \frac{\mathrm{d}^3 \bm{t}}{(2\pi)^3}  \left(\frac{t^2}{q|\bm{k}-\bm{q}|}\right)^\Delta P_\zeta(t) \,, \nonumber
\end{align}
where in the first line we recognize again the disconnected contribution to the tensor power spectrum, divided by two.
But the variable of integration $q$ appears in the second line, so we first compute this integral, which gives for a scale-invariant power spectrum, and defining $\tilde{t}=t/t_\star$:
\begin{equation}
    \frac{\tau_{\text{NL}}}{2} \mathcal{P}_\zeta
    \alpha^{2\Delta}
    \int_{1/kL}^{1} \mathrm{d}\tilde{t} \,
   \tilde{t}^{2\Delta-1}  = \frac{\tau_{\text{NL}}}{2} \mathcal{P}_\zeta \frac{\alpha^{2\Delta}}{2\Delta} \left[  1 - e^{-2\Delta \log kL}\right] \,,
\end{equation}
where we also introduced the finite comoving size $L$ as an IR cutoff for consistency.
Interestingly, this solution interpolates between a behavior $\propto \alpha^{2\Delta}/\Delta$ independent of the IR cutoff $L$ for a not-too-small $\Delta \gtrsim 1/\log kL$, and the massless case $\propto \log kL$ independent of $\alpha$.
The contribution is monotonously decreasing with increasing $\Delta$, i.e.\ the heavier the exchanged spin-$0$ particle, the smaller the signal.
The contribution from the $u$-channel seed can easily be shown to be the same, and using the expression of the total trispectrum-induced tensor power spectrum in terms of the seeds contributions, one therefore finds
\begin{tcolorbox}[colframe=white,arc=0pt,colback=greyish2]
\begin{equation}
\label{eq: result extended local shape}
    \frac{\Omega_{\mathrm{GW},\mathrm{c}}}{\Omega_{\mathrm{GW},\mathrm{d}}} \simeq  4\, \tau_{\text{NL}} \mathcal{P}_{\zeta}\, \frac{\alpha^{2\Delta}}{2\Delta} \left[  1 - e^{-2\Delta \log kL}\right].
\end{equation}
\end{tcolorbox}

\begin{figure}[t!]
   \centering
            \includegraphics[width=\linewidth]{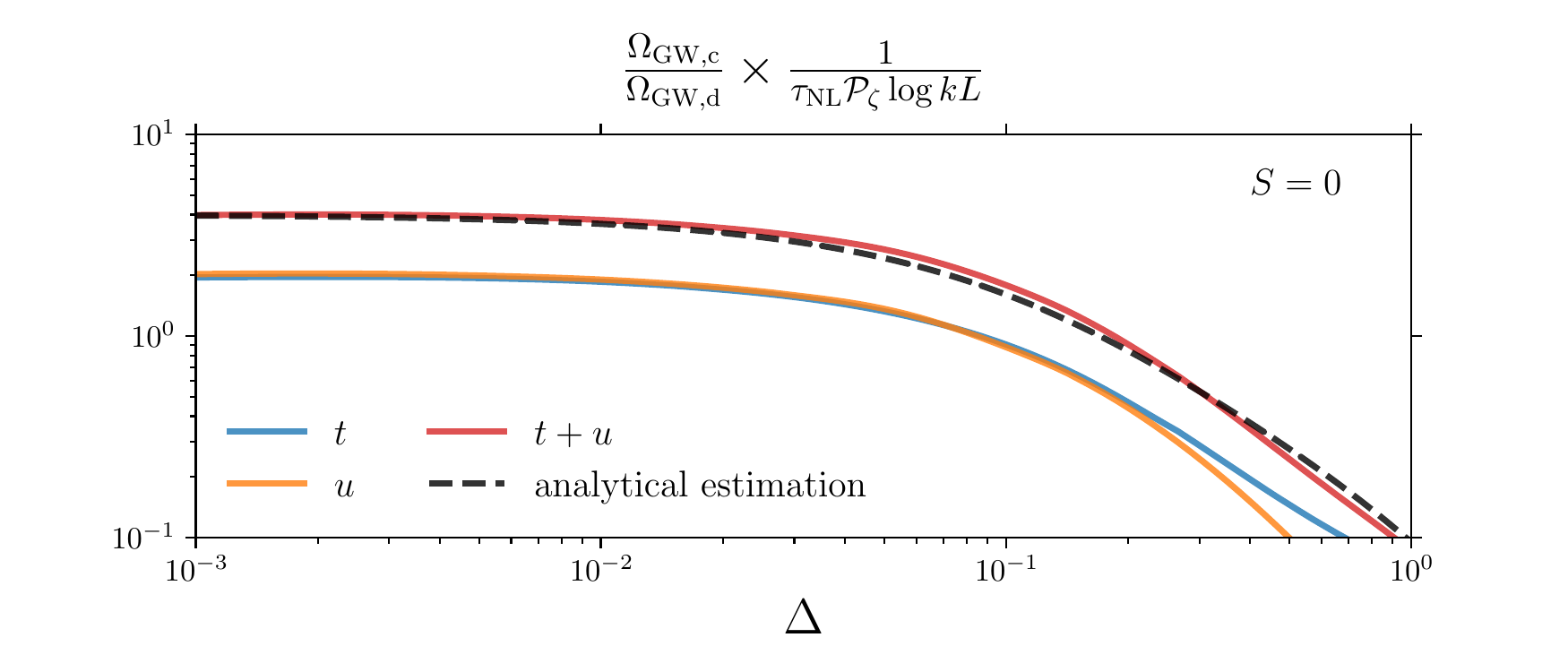}
   \caption{Connected (trispectrum-induced) contribution to the scalar-induced gravitational wave energy density, normalised by the disconnected contribution (Gaussian scalar fluctuations), as a function of $\Delta = 3/2 - \sqrt{9/4 - m^2/H^2}$, where $m$ is the mass of the non-spinning $S=0$ exchanged particle.
   The blue and orange lines correspond respectively to the contributions from the $t$- and the $u$-channel, and the total of the two is represented by the red line.The total signal asymptotes the analytical prediction $4$ of Eq.~\eqref{eq:ratio} in the limit of a massless spinless scalar field, and decreases as $1/\Delta$ for sizeable values of $\Delta$ as found in  Eq.~\eqref{eq: result extended local shape}.
    The black dashed line shows our analytical estimation of the signal for intermediate values, see Eq.~\eqref{eq: result extended local shape}, for $\alpha = 0.69$, which we find to be the best-fit value.
   }
  \label{fig: massive spin0}
\end{figure}

\noindent Without resorting to the assumptions that enabled one to derive the analytical expression of Eq.~\eqref{eq: result extended local shape}, one can numerically compute the integrals of the $t$ and $u$ channels contributions for any mass parameter $\Delta$ of the exchanged particle, and introducing the IR cutoff $L^{-1}$.
The result of the numerical integration is shown in Fig.~\ref{fig: massive spin0}, which confirms both the analytical behavior $1/\Delta$ away from the very light case, and the saturation of the signal by the IR cutoff when $\Delta \rightarrow 0$.
The overall shape of the signal closely resembles the analytical estimation for the intermediate regime if the parameter $\alpha$, quantifying the degree of collapsing in the analytical estimation, is chosen to be $\alpha = 0.69$, which is indeed consistent with the requirement $\alpha <1$.
See App.~\ref{app:numerics} for more details about the numerics.

\begin{figure}[h!]
   \centering
            \includegraphics[width=\linewidth]{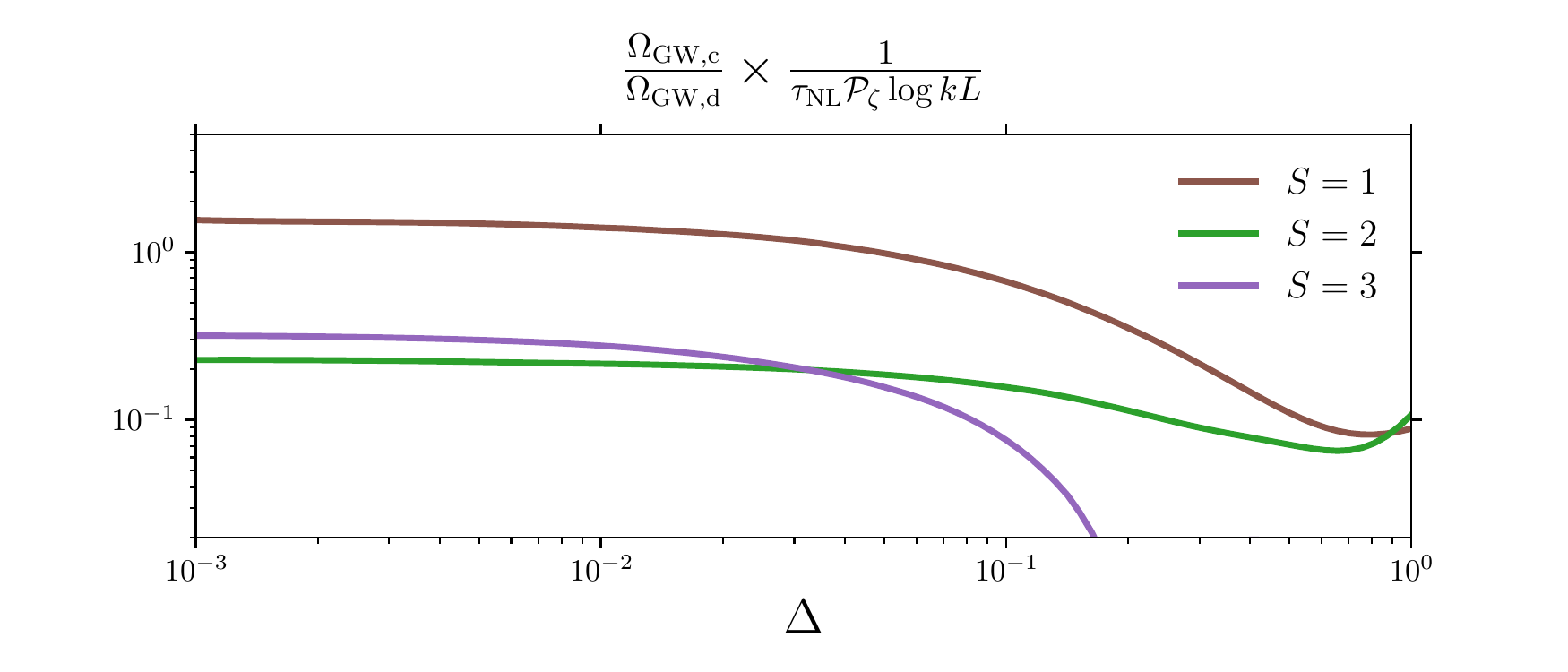}
   \caption{
   Connected contribution (trispectrum-induced) to the scalar-induced gravitational wave energy density, normalised by the disconnected contribution (Gaussian scalar fluctuations), as a function of $\Delta = 3/2 - \sqrt{9/4 - m^2/H^2}$, where $m$ is the mass of the exchanged particle, for $S=1$ (brown line), $S=2$ (green line) and $S=3$ (purple line).
   Each of these lines corresponds to the total contribution for a given spin, coming both from the $t$- and $u$-channels, as the $s$-channel contributions are always negligible, see the main text.}
  \label{fig: massive spin}
\end{figure}

\paragraph{Spinning and massive case.}
We now turn to the case of spinning and massive exchanged particles, for which the Legendre polynomials $P_S$ add an additional angular dependence to the integrand.
This angular dependence prevents us from deriving analytical results, so we simply resort to a numerical computation for spins $S=1,2,3$.
Moreover, because of this angular dependence, the $s$-channel now depends on the variables $t$ and $u$, and therefore gives a non-vanishing contribution to the gravitational wave energy density.
However, this contribution is always negligible compared to the ones from the $t$ and $u$ channels.\footnote{Indeed, remember that the regime of validity of the $s$-channel contribution corresponds to the collapsed configuration $s=|\bm{k}_1+\bm{k}_2| \ll k_a$.
From the master formula, Eq.~\eqref{eq: master formula vec(q) variables}, we read that $s=k$ the external momentum of the tensor modes, therefore the regime of validity of the $s$-channel corresponds to the UV region of integration in the multiple integrals: $q_i, |\bm{q}_i-\bm{k}| \gg k$.
In this region, the kernel of integration $\mathcal{K}_c$ is highly damped, see the discussion in section \ref{sec:kernels} and
App.~\ref{app:kernel}, which therefore suppresses the signal from the $s$-channel contribution.
}
The results of the numerical computation of the integrals are shown in Fig.~\ref{fig: massive spin}, see App.~\ref{app:numerics} for more details about the numerics.

\subsection{No-go theorem}
\label{sec: perturbativity bound}

The requirement of perturbativity or weak coupling has the implication that the 1-loop correction to the scalar power spectrum must be smaller than the tree-level result. This translates into a bound on the coupling constants of the theory and potentially other parameters describing primordial fluctuations. 
In this section, by identifying the strong coupling scales of the interactions leading to the trispectra of interest in the EFT of inflation, and by computing the 1-loop power spectrum within the $\delta N$ formalism in the case of the local shapes, we demonstrate that perturbative control necessarily results in the trispectrum-induced SGWB to be negligible compared to the one induced in the Gaussian approximation.

\subsubsection*{{\large Derivative interactions and regular shapes}}

\paragraph{Unitarity bound and weak coupling.}

In the EFT of inflation, the cutoff scale of the theory---beyond which scalar fluctuations become strongly coupled---can be determined by requiring that tree-level scattering amplitudes are unitary \cite{Baumann:2011su,Baumann:2014cja,Cheung:2007st}. One finds
\beq \label{eq:unitarity bound}
 \mathrm{unitarity} \quad \implies \quad E^4 \leqslant
\Lambda_u^4=\frac{6}{5 \pi}\frac{c_s^4}{1-c_s^2}\frac{H^4}{{\cal P}_\zeta} \,.
\eeq
Fluctuations with energy $E>\Lambda_u$ violate unitarity, therefore signaling the breakdown of the perturbative theory.We remark that this bound is actually derived in the particular case where the Wilson coefficients $A$ and $B$ of the EFT are set to zero (see the discussion around Eq.\ \eqref{eq:wilson coeffs} above) and need not be valid for all values of $A$ and $B$.

\vskip 4pt
On the other hand, a simpler cutoff scale can be found from the requirement of a weakly coupled dynamics which, in the regime of a small speed of sound of phenomenological interest for this work, is an equivalent statement up to order one numbers. One finds
\begin{equation}
    \mathrm{weak\,\,coupling\,\,of} \,\, \zeta^{\prime}(\partial\zeta)^2 \quad \implies \quad E \lesssim \Lambda_1 = H c_s \mathcal{P}_\zeta^{-1/4} \,,
\end{equation}
from which it is clear that $\Lambda_u \sim \Lambda_1$ when $c_s \ll 1$.
It is then straightforward to extend the weak coupling requirement to the other cubic interaction of the EFT of inflation involving the parameter $A$ which is therefore also bounded (from above):
\begin{equation}
    \mathrm{weak\,\,coupling\,\,of} \,\, \zeta^{\prime3} \quad \implies \quad E \lesssim \Lambda_2 = H c_s |A|^{-1/2} \mathcal{P}_\zeta^{-1/4} \,.
\end{equation}
For natural values of $A$ of order unity, the two cubic couplings display the same strength and the two strong coupling scales coincide.

\paragraph{No-go theorem.}

We may now express $\mathcal{P}_\zeta$ in terms of the strong coupling scales
$\Lambda_{1,2}$, which once
substituted in \eqref{eq: ratio regular shape} in terms of the EFT parameters $c_s^2$ and $A$, and
in the limit of a small speed of sound, gives
\beq
\frac{\Omega_{\mathrm{GW},\mathrm{c}}}{\Omega_{\mathrm{GW},\mathrm{d}}}=
\frac{\mathcal{I}_{s1}}{\mathcal{I}_{\rm d}} \left(\frac{H}{\Lambda_2}\right)^4 
-\frac{\mathcal{I}_{s2}}{\mathcal{I}_{\rm d}}\left(\frac{H}{\sqrt{\Lambda_1\Lambda_2}}\right)^4
+ \frac{\mathcal{I}_{s3}}{\mathcal{I}_{\rm d}}\, \left(\frac{H}{\Lambda_1}\right)^4  \,.
\eeq
Note that the term coming from the only contributing contact interaction, $\mathcal{I}_{c3}$, is suppressed by a factor of $c_s^2$ compared to the scalar-exchange ones, and was therefore omitted by consistency.
We also recall our result that each ratio $\mathcal{I}_{(i)}/\mathcal{I}_{\rm d}$ is small, of order $10^{-2}-10^{-1}$.
Requiring that the theory remains weakly coupled up to the Hubble scale at which primordial fluctuations exit the horizon, i.e.\ that it remains under perturbative control, amounts to asking that $H \lesssim \Lambda_\star \equiv \mathrm{min}\left(\Lambda_1,\Lambda_2\right)$.
This therefore severely bounds the relative size of the contribution from the trispectrum to the gravitational wave energy density, compared to the Gaussian-induced one:
\beq
\frac{\Omega_{\mathrm{GW},\mathrm{c}}}{\Omega_{\mathrm{GW},\mathrm{d}}}=\mathcal{O}(10^{-1})\left(\frac{H}{\Lambda_\star}\right)^4 \lesssim \mathcal{O}(10^{-1}) \,.
\eeq
This no-go theorem only relies on the requirement of weak coupling, and holds even for unnaturally large values of the parameter $A$.

\subsubsection*{\large{Massless fields and local interactions}}

\paragraph{Perturbativity bound.}

The calculation of the 1-loop scalar power spectrum in the $\delta N$ formalism was first performed by Lyth in \cite{Lyth:2007jh}. We next repeat this calculation generalized to an arbitrary number of degrees of freedom.

\vskip 4pt
The local ansatz up to quadratic order in the Gaussian fields $\sigma^I$ reads, in position space,
\beq \label{eq:local ansatz1}
\zeta=N_I\sigma^I+\frac{1}{2}\,N_{IJ}\sigma^I\sigma^J \,,
\eeq
where $N_I$ and $N_{IJ}$ are the (field-space covariant) derivatives of the local e-folding number $N$ with respect to the fields.
The real symmetric cross power spectra of the $\sigma^I$ at late times are defined via $\langle\sigma^I_{\bm{k}}\sigma^J_{\bm{k}'}\rangle=(2\pi)^3\delta(\bm{k}+\bm{k}')\Psig^{IJ}(k)$, so that the tree-level power spectrum of $\zeta$ is $P^{\rm(tree)}_{\zeta}=N_IN_J\Psig^{IJ}(k)$.
The 1-loop correction to the power spectrum is given by the 2-point function of the non-Gaussian term in \eqref{eq:local ansatz1}. To find this we use the identity $(\sigma^I\sigma^J)_{\bm{k}}=\int \frac{\mathrm{d}^3\bm{p}}{(2\pi)^3} \,\sigma^I_{\bm{p}}\sigma^J_{\bm{k}-\bm{p}}$
so that
\beq
N_{IJ}N_{KL}\langle(\sigma^I\sigma^J)_{\bm{k}}(\sigma^K\sigma^L)_{\bm{k}'}\rangle = 2\delta(\bm{k}+\bm{k}')N_{IJ}N_{KL}\int\mathrm{d}^3\bm{p}\,\Psig^{IK}(p)\Psig^{JL}(|\bm{k}-\bm{p}|) \,.
\eeq
The 1-loop power spectrum of $\zeta$ is then
\beq
P^{\rm(1-loop)}_{\zeta}=\frac{N_{IJ}N_{KL}}{2}\int \frac{\mathrm{d}^3\bm{p}}{(2\pi)^3} \,\Psig^{IK}(p)\Psig^{JL}(|\bm{k}-\bm{p}|) \,.
\eeq
Specifying to a scale-invariant $\sigma$-spectrum, we have $\Psig^{IJ}=(2\pi^2/k^3)\mathcal{P}^{IJ}$, where $\mathcal{P}^{IJ}$ is independent of $k$ but may in principle depend on various background quantities. In fact, in standard scenarios one has $\mathcal{P}^{IJ}\simeq (H/2\pi)^2G^{IJ}$ at horizon crossing, with $G^{IJ}$ the (inverse) field-space metric that defines the kinetic terms of the $\sigma^I$ fields. Since this metric must have Euclidean signature in order to avoid ghost degrees of freedom, we infer that $\mathcal{P}^{IJ}$ is positive-definite whenever the above relation holds. While we will keep $\mathcal{P}^{IJ}$ arbitrary for the sake of generality, we shall nevertheless make the reasonable
assumption that it is positive-definite. This is certainly very plausible on physical grounds, as can be seen by considering a basis where $\mathcal{P}^{IJ}$ is diagonal with the entries being simply the power spectra of the fields, which are positive in any standard scenario.

\vskip 4pt
In the scale-invariant case we then obtain
\beq\begin{aligned}
P^{\rm(1-loop)}_{\zeta} &= \frac{\pi}{4}\,N_{IJ}N_{KL}\mathcal{P}^{IK}\mathcal{P}^{JL}\int\mathrm{d}^3\bm{p}\,\frac{1}{p^3|\bm{k}-\bm{p}|^3} \\
&\simeq \frac{2\pi^2}{k^3}\,N_{IJ}N_{KL}\mathcal{P}^{IK}\mathcal{P}^{JL}\log(kL) \,,
\end{aligned}\eeq
where to get the second line we have regularized the IR-divergent integral with a momentum cutoff $L^{-1}$ and isolated the leading-order term.\footnote{We note that our result differs by a factor of 2 relative to \cite{Lyth:2007jh}, which seems to be a typo or oversight in the latter reference.} As usual, the details of the regularization scheme do not matter in the limit of large $kL$, so this is guaranteed to be the dominant contribution to the integral whenever the logarithm dominates. We thus obtain the following result for the perturbativity bound at 1-loop:
\beq \label{eq:pert bound local}
P^{\rm(1-loop)}_{\zeta}/P^{\rm(tree)}_{\zeta}=\frac{N_{IJ}N_{KL}\mathcal{P}^{IK}\mathcal{P}^{JL}}{N_IN_J\mathcal{P}^{IJ}}\,\log(kL) < 1 \,.
\eeq

\paragraph{No-go theorem.}

In order to relate the previous bound to the ratio $\Omega_{\mathrm{GW},\mathrm{c}}/\Omega_{\mathrm{GW},\mathrm{d}}$ we derived above in \eqref{eq:ratio}, we recall that in the $\delta N$ formalism the parameter $\tau_{\rm NL}$ can be expressed as~\cite{Byrnes:2006vq}
\beq
\tau_{\rm NL}=\frac{N_IN_JN_{KL}N_{MN}\mathcal{P}^{IK}\mathcal{P}^{JM}\mathcal{P}^{LN}}{(N_IN_J\mathcal{P}^{IJ})^3} \,.
\eeq
Therefore, in the case of a single degree of freedom, we immediately obtain
\beq
\tau_{\rm NL}\mathcal{P}_{\zeta}\log(kL) = P^{\rm(1-loop)}_{\zeta}/P^{\rm(tree)}_{\zeta} < 1 \,,
\eeq
and so $\Omega_{\mathrm{GW},\mathrm{c}}/\Omega_{\mathrm{GW},\mathrm{d}} < 4$. Of course, to be on the conservative side, one would usually require the ratio $P^{\rm(1-loop)}_{\zeta}/P^{\rm(tree)}_{\zeta}$ to be parametrically small, and hence we infer the condition
\beq
\frac{\Omega_{\mathrm{GW},\mathrm{c}}}{\Omega_{\mathrm{GW},\mathrm{d}}} \ll 1 \,,
\eeq
in order for the theory to be weakly coupled. This is our no-go result on the relative contribution of the trispectrum to $\Omega_{\rm GW}$.

\vskip 4pt
The question is then whether a model with multiple fields can in principle evade this outcome. The answer is no, since in general we have
\beq
\tau_{\rm NL}\mathcal{P}_{\zeta}\log(kL) \leqslant P^{\rm(1-loop)}_{\zeta}/P^{\rm(tree)}_{\zeta} \,,
\eeq
where the equality holds only in the case of a single field.

\vskip 4pt
For the proof, let us first observe that the previous inequality may be equivalently written as
\beq
\frac{\vec{N}^T\bm{N}^T\bm{\mathcal{P}N}\vec{N}}{\vec{N}^T\bm{\mathcal{P}}\vec{N}} \leqslant {\rm tr}(\bm{N}^2) \,,
\eeq
where $\bm{N}$ is the matrix with entries $N^I_{\phantom{I}J}\equiv\mathcal{P}^{IK}N_{KJ}$, $\vec{N}$ is the vector with entries $N^I\equiv\mathcal{P}^{IJ}N_J$, and $\bm {\mathcal{P}}$ is the matrix with entries $\mathcal{P}_{IJ}$. Note that $\bm{N}$ is generically \textit{not} symmetric. However, its eigenvalues are real, as the following lemma shows.

\begin{framed}
\textbf{Lemma.} Given real symmetric matrices $\bm{A}$ and $\bm{B}$, with $\bm{A}$ positive definite, then $\bm{C}\equiv \bm{AB}$ is diagonalizable (over the complex numbers) and has real eigenvalues.

\vskip 4pt
To prove that $\bm{C}$ is diagonalizable, observe that $\bm{A}$ possesses an invertible symmetric square root $\bm{A}^{1/2}$. Hence,
\beq
\bm{A}^{-1/2}\bm{CA}^{1/2}=\bm{A}^{1/2}\bm{BA}^{1/2} \,,
\eeq
which shows that $\bm{C}$ is similar to a symmetric matrix and therefore diagonalizable.
For the second proposition, consider the eigenvalue equations for $\bm{B}$ and $\bm{C}$,
\beq
\bm{B}\vec{b}=b\vec{b} \,,\qquad \bm{C}\vec{c}=c\vec{c} \,.
\eeq
Since they are diagonalizable, we may choose each set of eigenvectors to be orthonormal (with respect to the complex Euclidean norm, although the $\vec{b}$'s may be chosen as real). We may then express the identity matrix as $\bm{1}=\sum_b\vec{b}\,\vec{b}^{\,T}=\sum_c\vec{c}\,\vec{c}^{\,*T}$, so that
\beq
c=\vec{c}^{\,*T}\bm{AB}\vec{c}=\sum_{b}b(\vec{c}^{\,*T}\bm{A}\vec{b})(\vec{b}^{\,T}\vec{c})=\sum_{b,c}b(\vec{c}^{\,*T}\bm{A}\vec{c})|(\vec{b}^{\,T}\vec{c})|^2 \,.
\eeq
But by assumption $b$ is real and $\vec{c}^{\,*T}\bm{A}\vec{c}>0$, hence $c$ is real.
\end{framed}

\noindent Consider the invertible symmetric square root matrix of $\bm{\mathcal{P}}$, $\bm{\mathcal{P}}^{1/2}$, and note that $\bm{\mathcal{P}}^{-1}\bm{N}^T\bm{\mathcal{P}N}=\bm{N}^2$ is positive semi-definite from the lemma (but not symmetric).
Therefore, inserting twice the identity matrix $\bm{1}=\bm{\mathcal{P}}\bm{\mathcal{P}}^{-1}=\bm{\mathcal{P}}^{-1/2}\bm{\mathcal{P}}^{1/2}$ one has:
\beq
\frac{\vec{N}^T\bm{N}^T\bm{\mathcal{P}N}\vec{N}}{\vec{N}^T\bm{\mathcal{P}}\vec{N}} = \frac{(\bm{\mathcal{P}}^{1/2}\vec{N})^T\bm{\mathcal{P}}^{1/2}\bm{N}^2\bm{\mathcal{P}}^{-1/2}(\bm{\mathcal{P}}^{1/2}\vec{N})}{(\bm{\mathcal{P}}^{1/2}\vec{N})^T(\bm{\mathcal{P}}^{1/2}\vec{N})} \equiv R(\bm{M},\vec{X})  \,,
\eeq
which is the Rayleigh-Ritz quotient $R(\bm{M},\vec{X})$ of the symmetric positive semi-definite matrix $\bm{M}=\bm{\mathcal{P}}^{1/2}\bm{N}^2\bm{\mathcal{P}}^{-1/2}$ 
(indeed $\bm{M}^T=(\bm{\mathcal{P}}^{-1/2}\bm{N}^T\bm{\mathcal{P}N}\bm{\mathcal{P}}^{-1/2})^T=\bm{M}$)
and the vector $\vec{X}=\bm{\mathcal{P}}^{1/2}\vec{N}$.
Denoting $0\leqslant\nu\leqslant \nu_\mathrm{max}$ the eigenvalues of $\bm{M}$, one has that $R(\bm{M},\vec{X})\leqslant\nu_\mathrm{max}$ with equality if and only if $\vec{X}$ is proportional to $\vec{X}_\mathrm{max}$ the eigenvector of $M$ associated to the largest eigenvalue $\nu_\mathrm{max}$.
Moreover, we have trivially that:
\beq
\nu_{\rm max}\leqslant \sum_{\nu}\nu={\rm tr}(\bm{M})= {\rm tr}(\bm{N}^2) \,,
\eeq
with equality iff the dimension of the field space is 1, or if all $\nu$ but $\nu_\mathrm{max}$ are vanishing.
This completes the proof.

\section{Discussion}

The aim of this paper was to perform a first assessment of the importance of the trispectrum to the GW background induced by primordial scalar perturbations. Unlike previous works, we have assumed what is perhaps the most conservative and agnostic scenario in which scale invariance is unbroken. Even within this restricted set-up, we have not analyzed all the possible trispectrum shapes, but rather focused on the two classes that are best motivated theoretically, namely the ``equilateral'' shapes predicted by the low-energy EFT of inflationary fluctuations and the local-type shapes that arise in models with multiple light fields.

\vskip 4pt
Our main result can be stated as a no-go theorem, as we have shown that the ratio $\Omega_{\mathrm{GW},\mathrm{c}}/\Omega_{\mathrm{GW},\mathrm{d}}$ for the connected and disconnected contributions to $\Omega_{\mathrm{GW}}$ is in every case bounded by a small number from the requirement of having a consistent perturbative description. In more detail, we find the relation $\Omega_{\mathrm{GW},\mathrm{c}}/\Omega_{\mathrm{GW},\mathrm{d}}=(\mathcal{I}_{\rm c}/\mathcal{I}_{\rm d})(H/\Lambda_{\star})^4$ for the regular shapes that follow from the standard derivative interactions, where $\mathcal{I}_{\rm c}/\mathcal{I}_{\rm d}=\mathcal{O}(10^{-2}-10^{-1})$ depending on the specific shape, and $\Lambda_{\star}$ is the strong-coupling scale of the theory. The contribution of the trispectrum is therefore necessarily subleading, even if one were to push the Hubble scale to values comparable to $\Lambda_{\star}$.

\vskip 4pt
The situation is similar in the context of local NGs, as we find the inequality $\Omega_{\mathrm{GW},\mathrm{c}}/\Omega_{\mathrm{GW},\mathrm{d}}\leqslant 4(P^{\rm(1-loop)}_{\zeta}/P^{\rm(tree)}_{\zeta})$. Here the situation is slightly more promising because of the factor of 4, but it should be remembered that generically the inequality is strict. Moreover, if one wishes to remain on the conservative side of weak coupling, one should not allow $P^{\rm(1-loop)}_{\zeta}/P^{\rm(tree)}_{\zeta}$ to reach values close to unity. It would be interesting to scrutinize in more detail the situation where loop corrections are sizeable, say $P^{\rm(1-loop)}_{\zeta}/P^{\rm(tree)}_{\zeta}\sim 10^{-1}$, but still within perturbative control. Another related open issue concerns the validity of our no-go result in the case of local interactions with massive mediators with or without spin. The technical challenge to overcome is the calculation of $P^{\rm(1-loop)}_{\zeta}$ and hence the corresponding perturbativity bound. Although we have seen that $\Omega_{\mathrm{GW},\mathrm{c}}/\Omega_{\mathrm{GW},\mathrm{d}}$ decreases with the mass of the mediator, the ratio $P^{\rm(1-loop)}_{\zeta}/P^{\rm(tree)}_{\zeta}$ should similarly be reduced as loop corrections become suppressed for massive fields. Thus the naive expectation is that the final conclusion will be essentially unchanged, although a careful calculation would certainly be worthwhile.

\vskip 4pt
The natural next step would be to explore if our no-go result will persist upon breaking the perfect scale invariance we have assumed. Here we conclude our paper by carrying out a preliminary study of this question using a simple toy model and focusing on local-type NGs, leaving a more general analysis to future work. The simplest way in which the scale invariance of the scalar power spectrum may be broken is by the existence of a sharp step, raising the amplitude of $\mathcal{P}_\zeta(k)$ on small scales to potentially observable values.

\vskip 4pt
We model this step feature as
\begin{equation}
\mathcal{P}_\zeta(k)=A_s\left[1+\alpha\Theta(k-k_0)\right] \,,
\end{equation}
where $A_s$ is the amplitude of the spectrum on CMB scales, which gets boosted in this model to $A_s(1+\alpha)\simeq \alpha A_s$, assuming that $\alpha\gg1$ having in mind the possibility of a detectable GW signal on small scales. A second assumption we make is that $k\gg k_0$ for the scale $k$ of interest, i.e.\ we assume the step occurs at relatively large scales, parametrically between the scales probed in the CMB and those of observational relevance for interferometers.

\vskip 4pt
These two assumptions, namely $\alpha\gg1$ and $k\gg k_0$, greatly simplify the calculation of the GW spectrum. Indeed, the structure of the local ansatz, Eq.\ \eqref{eq: local shape}, shows that the result for $P^{\text{c}}_h$ will be a cubic polynomial in $\alpha$, and our assumption then implies that we can isolate the term proportional to $\alpha^3$. But this term is nothing but the GW spectrum we have already computed in the exactly scale-invariant case, only that the integrals are now cut off by the scale $k_0$ instead of the IR cutoff $L^{-1}$. Similarly, for the disconnected 4-point function, we have the term proportional to $\alpha^2$ as the dominant contribution, and since the master integral is convergent in the case of exact scale invariance, we know that the cutoff $k_0$ will not appear in the leading-order result due to the assumption $k\gg k_0$. We therefore conclude that, on scales $k\gg k_0$,
\begin{equation}
\Omega_{\mathrm{GW},\mathrm{c}}/\Omega_{\mathrm{GW},\mathrm{d}} \simeq 4\alpha A_s\log(k/k_0) \,.
\end{equation}
The unsurprising result is then that the ratio $\Omega_{\mathrm{GW},\mathrm{c}}/\Omega_{\mathrm{GW},\mathrm{d}}$ takes the same form as in the scale-invariant set-up, only with the scalar spectrum amplitude evaluated on small scales and with the role of the IR cutoff now played by $k_0$. The reader may be concerned that we have neglected the ``divergent'' term proportional to $\alpha^2\log(kL)$ in ${P^{\text{c}}_h}$ in comparison with the manifestly finite one proportional to $\alpha^3$. This approximation thus requires $\alpha\gg \log(kL)$, which is not at all unrealistic. For instance, having the LISA experiment in mind, we may estimate $k\sim 10^{12}\,{\rm Mpc^{-1}}$, $L^{-1}\sim 10^{-3}\,{\rm Mpc^{-1}}$ and $\alpha\sim 10^5$ (so as to have a potentially observable amplitude $\alpha A_s\sim 10^4$), so that the condition is clearly satisfied.

\vskip 4pt
The same considerations hold for the calculation of the 1-loop scalar power spectrum and the perturbativity bound derived from it. In this case, $P_{\zeta}^{\rm (1-loop)}$ has a leading contribution proportional to $\alpha^2$ and a loop integral cut off by $k_0$. Hence we readily infer, cf.\ \eqref{eq:pert bound local},
\begin{equation}
    P_{\zeta}^{\rm (1-loop)}/P_{\zeta}^{\rm (tree)} \simeq \alpha A_s\log(k/k_0) \,,
\end{equation}
assuming $\alpha\gg\log(kL)$ and $k\gg k_0$ as before. The clear implication is that the relative contribution of the connected 4-point function to the GW spectrum is of the same order as the perturbativity bound, so that our conclusions drawn from the scale-invariant set-up hold also in this case.

\vskip 4pt
It would surely be premature to suggest the existence of a general no-go theorem based on this simple scale-dependent model. Nevertheless, together with the main results of the paper, it is an intriguing outcome that motivates a more general study.

\vspace{0.5cm}
\paragraph{Acknowledgements.} 
We are grateful to Matteo Braglia, Guillem Domenech, Jacopo Fumagalli, Sadra Jazayeri, Caner Unal and Lukas T.~Witkowski for interesting discussions related to the content of this paper. We also thank the participants of the \href{https://indico.in2p3.fr/event/23850/}{Gravitational-Wave Primordial Cosmology Workshop}, held remotely at IAP, from which this project was initiated.
LP would like to acknowledge support from the “Atracci\'{o}n de Talento” grant 2019-T1/TIC15784, his work is partially supported by the Spanish Research Agency (Agencia Estatal de Investigaci\'{o}n) through the Grant IFT Centro de Excelencia Severo Ochoa No CEX2020-001007-S, funded by MCIN/AEI/10.13039/501100011033. SRP and DW are supported by the European Research Council under the European Union’s Horizon 2020 research and innovation programme (grant agreement No 758792, Starting Grant project GEODESI).

\newpage
\appendix

\section{Details on numerics}
\label{app:numerics}
%
%

In this appendix, we present details of the numerical computations of Section \ref{sec:pheno}. In particular, we will give explicit expressions of the recasted numerical integrals in Eq.\ (\ref{eq: master formula uv variables}).

\paragraph{Integral recast.} We are interested in individual channels of the polarization-summed gravitational wave power spectrum defined in terms of the trispectrum of curvature perturbation via the master integral (\ref{eq: master formula uv variables}). Given the expression of the transfer function (\ref{eq:Isquared avg}) and momentum-dependence of the trispectrum, it is convenient to define the following variables

\begin{equation}
\label{eq:def_u_v}
    u_i = \frac{|\bm{k} - \bm{q}_i|}{k}, \hspace*{0.5cm} v_i = \frac{q_i}{k}.
\end{equation}
The integration measure becomes

\begin{equation}
    \int \mathrm{d}^3q_i = k^3\int_0^{\infty}\mathrm{d}v_i\,v_i\int_{|1-v_i|}^{1+v_i}\mathrm{d}u_i\,u_i\int_0^{2\pi}\mathrm{d}\phi_i \,.
\end{equation}
Using these variables, the projection factor reads

\begin{equation}
Q_{\lambda}(\bm{k},\bm{q}_i)=\frac{k^2}{4\sqrt{2}}\left(4v_i^2-(1+v_i^2-u_i^2)^2\right)\alpha_{\lambda}(\phi_i) \,,
\end{equation}
where
\begin{equation}
\alpha_{\lambda}(\phi_i):= \begin{cases}
\cos2\phi_i & \mbox{if $\lambda=+$} \\
\sin2\phi_i & \mbox{if $\lambda=\times$}. \\
\end{cases}
\end{equation}
For numerical purposes, it is more suitable to integrate over a rectangular domain. To this end, we perform the following change of variables 

\begin{equation}
    u_i = \frac{t_i + s_i+1}{2}, \hspace*{0.5cm} v_i = \frac{t_i - s_i+1}{2}.
\end{equation}
We stress that the $t_i$ and $s_i$ variables are not the Mandelstam-like variables introduced in Section \ref{subsec:Dissecting the trispectrum}. The integration measure becomes

\begin{equation}
    \int \mathrm{d}^3q_i = \frac{k^3}{2}\int_0^{\infty}\mathrm{d}t_i\,v_i\int_{-1}^{+1}\mathrm{d}s_i\,u_i\int_0^{2\pi}\mathrm{d}\phi_i \,.
\end{equation}

\paragraph{Trispectrum momentum dependence.} The trispectrum $T_\zeta$ is defined by

\begin{equation}
    \braket{\zeta_{\bm{k}_1} \zeta_{\bm{k}_2} \zeta_{\bm{k}_3} \zeta_{\bm{k}_4}} = (2\pi)^3 \delta(\bm{k}_1+\bm{k}_2+\bm{k}_3+\bm{k}_4)\, T_\zeta (\bm{k}_1, \bm{k}_2, \bm{k}_3, \bm{k}_4).
\end{equation}
Note that inside the integral (\ref{eq: master formula vec(q) variables}), the arguments of the trispectrum correspond to 

\begin{equation}
    \bm{k}_1 = \bm{q}_1, \hspace*{0.5cm} \bm{k}_2 = \bm{k} - \bm{q}_1, \hspace*{0.5cm} \bm{k}_3 = -\bm{q}_2, \hspace*{0.5cm} \bm{k}_4 = -\bm{k} + \bm{q}_2,
\end{equation}
which makes it explicit that the $u_i$ and $v_i$ variables correspond to the norm of the trispectrum external moments $k_i$ in units of $k$

\begin{equation}
    v_1 = \frac{k_1}{k}, \hspace*{0.5cm} u_1 = \frac{k_2}{k}, \hspace*{0.5cm} v_2 = \frac{k_3}{k}, \hspace*{0.5cm} u_2 = \frac{k_4}{k}.
\end{equation}
For completeness, we also give the norm of internal momenta in terms of the integration variables. They read

\begin{equation}
\begin{aligned}
    |\bm{k}_1 + \bm{k}_2|  &= k \hspace*{0.5cm} &\text{($s$-channel)}, \\
    |\bm{k}_1+\bm{k}_3| &= k \, \sqrt{v_1^2 + v_2^2 - 2\,\bm{q}_1\cdot\bm{q}_2}\hspace*{0.5cm} &\text{($t$-channel)},\\
    |\bm{k}_2+\bm{k}_3| &= k \, \sqrt{1+v_1^2+v_2^2 - 2\,\bm{k}\cdot (\bm{q}_1+\bm{q}_2) + 2\,\bm{q}_1\cdot \bm{q}_2}\hspace*{0.5cm} &\text{($u$-channel)},
\end{aligned}
\end{equation}
where the dot products are given by

\begin{equation}
\begin{aligned}
    \frac{\bm{k}\cdot\bm{q}_i}{k^2} &= \frac{1}{2}\left(1+v_i^2-u_i^2\right), \\
    \frac{\bm{q}_1\cdot\bm{q}_2}{k^2} &= \frac{\cos(\phi_1-\phi_2)}{4}\sqrt{\left(4v_1^2-(1+v_1^2-u_1^2)^2\right)\left(4v_2^2-(1+v_2^2-u_2^2)^2\right)} \\
&\quad +\frac{1}{4}\left(1+v_1^2-u_1^2\right)\left(1+v_2^2-u_2^2\right).
\end{aligned}
\end{equation}
In the case of exchanged spinning fields, the trispectrum acquires an additional angular dependence

\begin{equation}
    \hat{\bm{k}}_1\cdot \hat{\bm{k}}_3 \hspace*{0.5cm} \text{($s$-channel)}, \hspace*{0.5cm}
    \hat{\bm{k}}_1\cdot \hat{\bm{k}}_2 \hspace*{0.5cm} \text{($t$-channel)}, \hspace*{0.5cm}
    \hat{\bm{k}}_2\cdot \hat{\bm{k}}_3 \hspace*{0.5cm} \text{($u$-channel)}.
\end{equation}
Because we live in a finite observable universe of characteristic size $H_0^{-1}$, modes with wavelength larger than a maximum size $L$ of a few Hubble radii are not described as perturbations but as part of the background. Therefore, one must impose a cosmological cutoff on the internal momentum integrals that formally diverge. We do so by applying a Heaviside function $\Theta(k_{ij} - L^{-1})$, where $k_{ij}$ is the internal momentum that depends on the considered channel. For numerical applications, we take $\log kL = 10$. We note that for a mode $k$ of observational relevance, say for LISA, the hierarchy $k L$ is much larger, but is more difficult to deal with numerically. Our choice is not problematic though because once the effect of a sufficient hierarchy like $\log kL = 10$ is taken into account, changing from such IR cutoff to a larger one can easily be taken into account, as the cutoff dependence is known analytically. We use the Monte Carlo estimator \textsf{Vegas} \cite{PETERLEPAGE1978192} to perform the multi-dimensional integrals with $30$ iterations of the vegas algorithm, each of which uses $10^7$ points to evaluate the integrand. These parameters enables us to achieve a relative precision of $10^{-3}$.

\section{Kernel}
\label{app:kernel}

In this appendix we investigate the structure of the kernel of the connected contribution in Eq.~\eqref{eq:kernel}, which is important in order to understand the graviton production.
Specifically, we show that it always selects modes around $k$---the produced graviton momentum---, and that it regulates the master integral (\ref{eq: master formula uv variables}) both in the IR and in the UV.

\begin{figure}[h!]
   \centering
            \includegraphics[width=\linewidth]{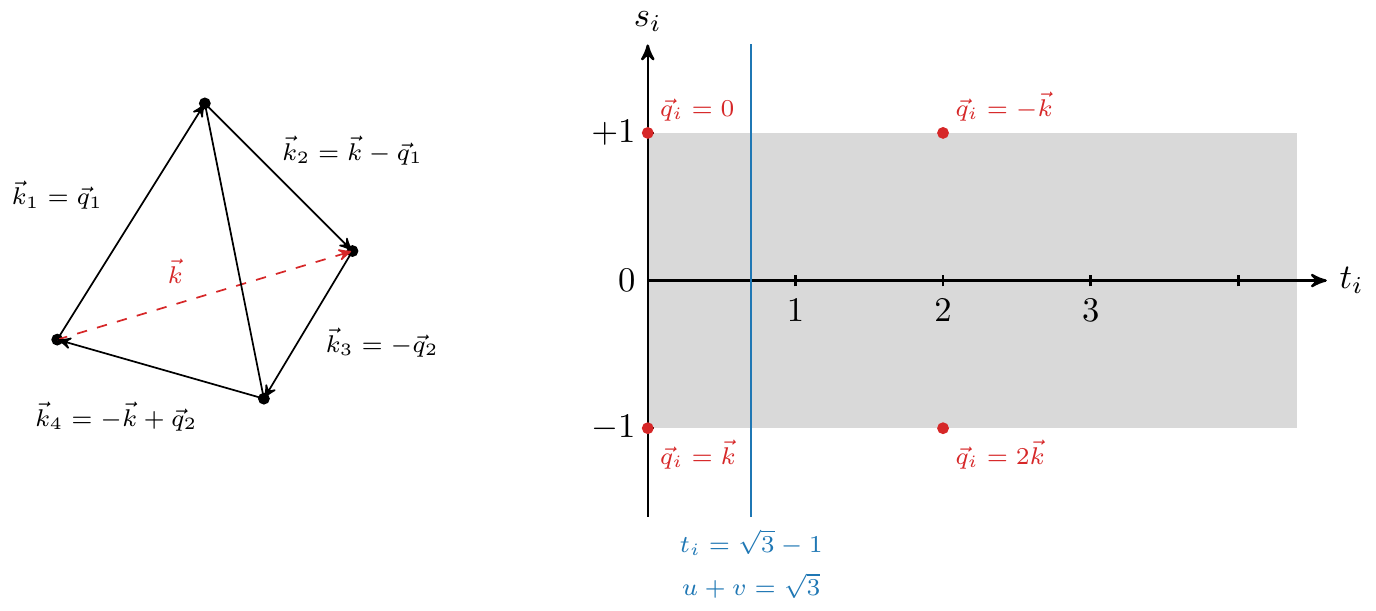}
   \caption{\textit{Left}: Representation of the trispectrum tetrahedron as it is parameterised inside the integral (\ref{eq: master formula vec(q) variables}) in terms of the variables $\bm{q}_i$. \textit{Right}: Domain of integration in the $(s_i, t_i)$ variables. We have depicted with red dots some characteristic tetrahedron configurations, and have denoted in blue the radiation domination resonance appearing in the kernel.}
  \label{fig: t-s plane}
\end{figure}

\begin{figure}[h!]
   \centering
     \includegraphics[width=\linewidth]{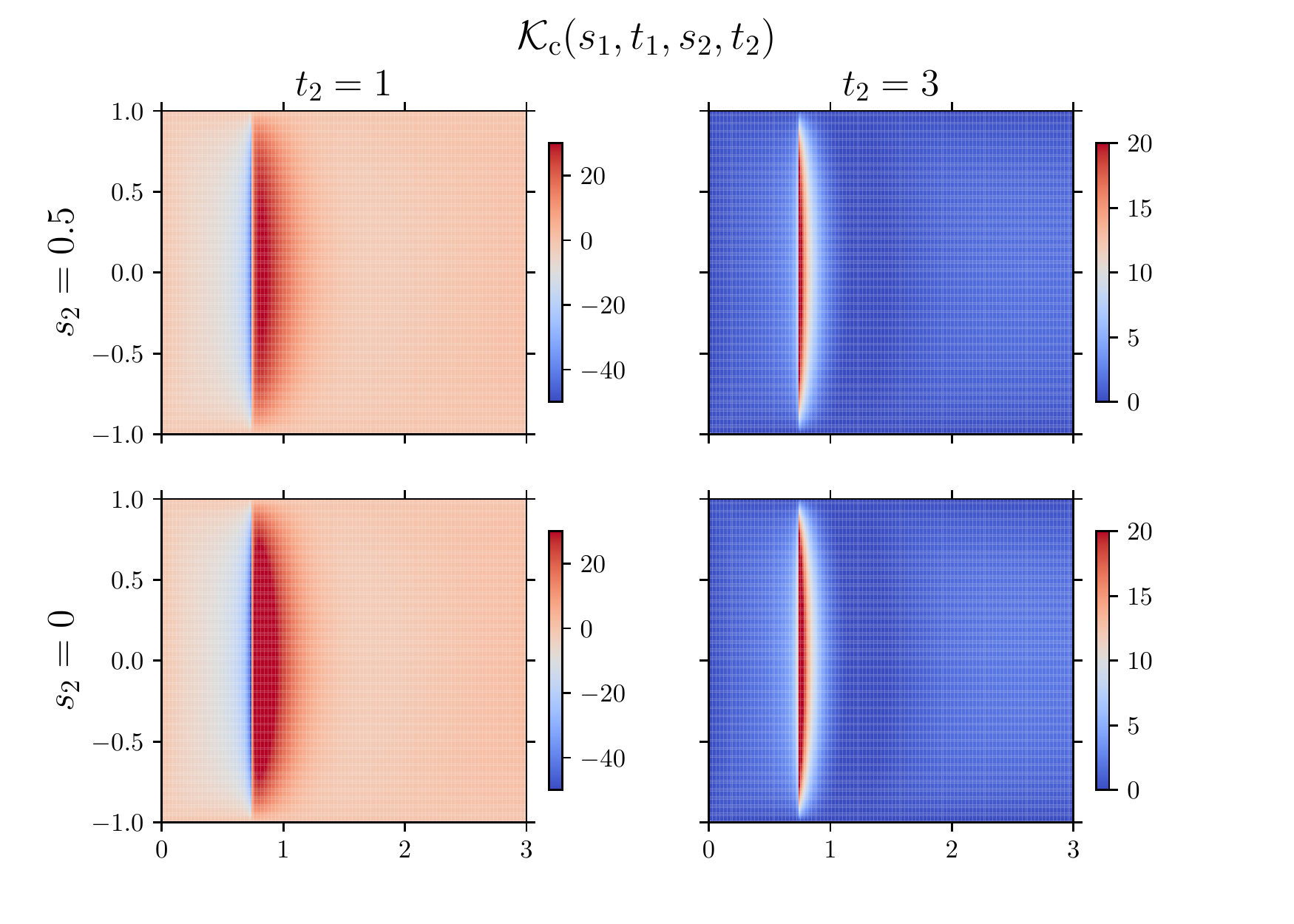}
   \caption{The kernel $\mathcal{K}_\mathrm{c}$ of the connected contribution in the $(s_1, t_1)$ plane for $s_2\in\{0, 0.5\}$ and $t_2\in\{1, 3\}$. Note that the kernel is symmetric under $s_i\rightarrow -s_i$, which can be checked explicitly.}
  \label{fig: kernel}
\end{figure}

\begin{figure}[h!]
   \centering
            \includegraphics[width=\linewidth]{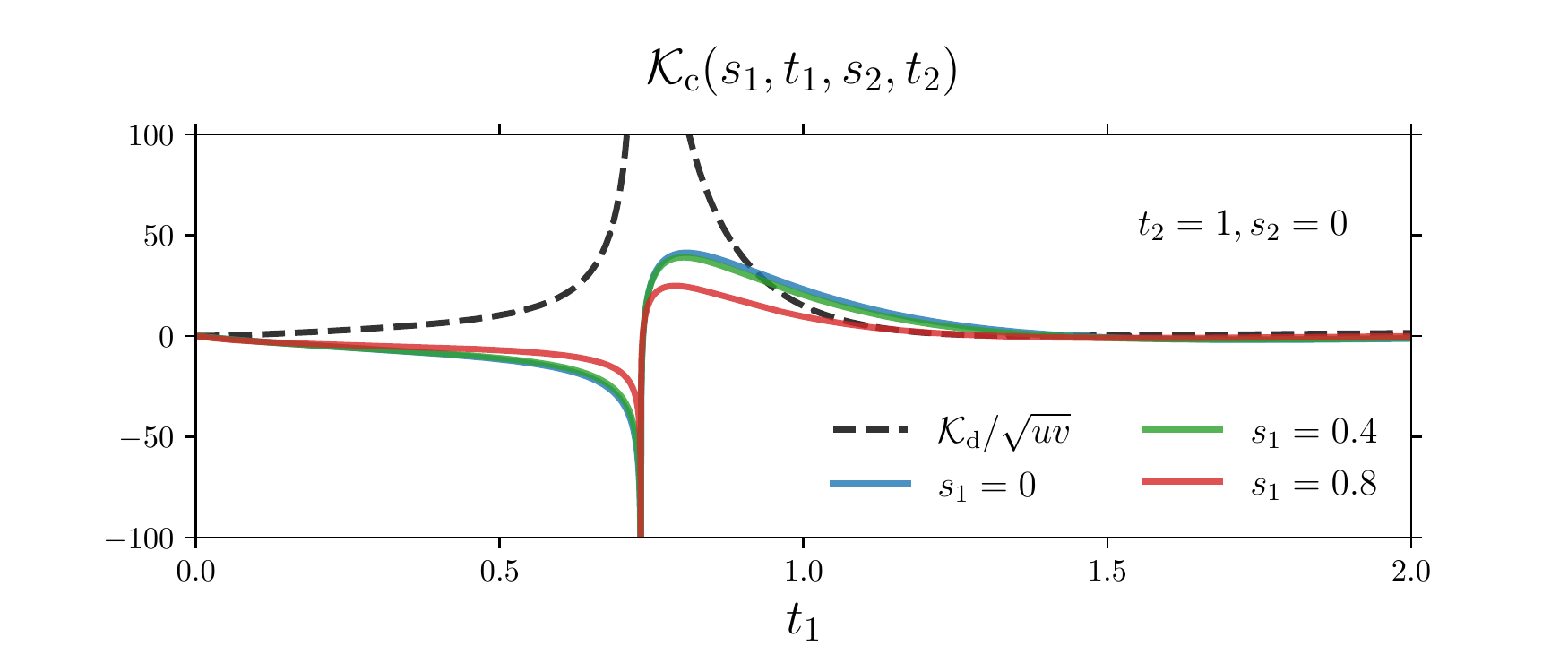}
   \caption{The kernel $\mathcal{K}_\mathrm{c}$ of the connected contribution as a function of $t_1$ and for a few values of $s_1\in\{0, 0.4, 0.8\}$, for fixed $(t_2, s_2)=(1,0)$.
   It vanishes as $\mathcal{K}_\mathrm{c}  \underset{t_1 \rightarrow 0}{\sim}  t_1$ in the IR and as $\mathcal{K}_\mathrm{c}  \underset{t_1 \rightarrow \infty}{\sim} \mathrm{log}(t_1)\,t_1^{-5/2}$ in the UV. For comparison, the dashed black line corresponds to the rescaled kernel $\mathcal{K}_{\mathrm{d}}/\sqrt{uv}$ of the disconnected contribution for $s=0.5$. We recall that $\mathcal{K}_{\mathrm{c}}(u, v, u, v) = \mathcal{K}_{\mathrm{d}}(u, v)/\sqrt{uv}$.
   }
  \label{fig:kernel_slice}
\end{figure}

\vskip 4pt
Let us first examine the structure of the kernel. First we note that the system is described by five vectors, one from the tensor mode $\bm{k}$ and four scalar modes $\bm{k}_1 = \bm{q}_1$, $\bm{k}_2 = \bm{k} - \bm{q}_1$, $\bm{k}_3 = - \bm{q}_2$ and $\bm{k}_4 = -\bm{k}+\bm{q}_2$, which of course satisfy momentum conservation $\sum_i\bm{k}_i = 0$. The scalar mode momenta form a tetrahedron that is depicted on the left panel of Fig. \ref{fig: t-s plane}. We also give some characteristic tetrahedron configurations that correspond to specific points---shown in red---on the right panel of Fig. \ref{fig: t-s plane}. We also have momentum conservation from two scalar modes producing a graviton mode. This means that these three variables satisfy a triangle inequality, which in terms of the $u_i, v_i$ variables defined in Eq. (\ref{eq:def_u_v}), reads

\begin{equation}
\label{eq:triangle_inequality}
    |1-v_i|<u_i<1+v_i \hspace*{0.5cm} \text{or} \hspace*{0.5cm} |u_i-v_i|<1<u_i+v_i\,.
\end{equation}
The kernel (\ref{eq:kernel}) is shown in Fig. \ref{fig: kernel} in the $(s_1, t_1)$ plane. One can notice the resonance at $t_1 = \sqrt{3} - 1$, equivalently $u_i+v_i=\sqrt{3}$, in the $I_B$ function entering the kernel, that comes from the radiation era $w=1/3$. This divergence does not saturate the triangle inequality (\ref{eq:triangle_inequality}). In the IR, the kernel scales as

\begin{equation}
    \mathcal{K}_{\text{c}} \sim t_1 t_2\,,
\end{equation}
which makes the integral converging once the kernel is multiplied by the trispectrum. In the UV, the kernel scales as 

\begin{equation}
    \mathcal{K}_{\text{c}} \sim \frac{\log(t_1) \log(t_2)}{t_1^{5/2} t_2^{5/2}}\,.
\end{equation}
This behaviour makes the integral converge in the UV. Therefore, as argued in Section \ref{subsec: extended local shapes}, the $s-$channel contribution -- being dominated by the UV part of the integral -- is negligible. We display in Fig. \ref{fig:kernel_slice} slices of the kernel in the $(s_1, t_1)$ plane.

\clearpage
\phantomsection
\addcontentsline{toc}{section}{References}
\bibliographystyle{utphys}
{\linespread{1.075}
\bibliography{references.bib}
}

\end{document}